\documentclass[prd,amsmath,aps,floats,amssymb, floatfix,
  superscriptaddress,nofootinbib,twocolumn,preprintnumbers]{revtex4}
\usepackage{tabularx}
\usepackage{amsmath,amssymb,natbib,latexsym,times}

\newcounter{blind_id}
\usepackage{graphicx}% Include figure files
\usepackage{dcolumn}% Align table columns on decimal point
\usepackage{bm}% bold math
\usepackage{url}
\usepackage{color}
\usepackage{comment}
\usepackage{mathrsfs}
\usepackage{hyperref}% add hypertext capabilities
\usepackage[dvipsnames]{xcolor}
\usepackage{orcidlink}

\newcommand{\simgt}{\lower.5ex\hbox{$\; \buildrel > \over \sim \;$}}

\newcommand{\bhat}[1]{{\hat{\bf {#1}}}}

\begin{document}

\title{On the equivalence between
galaxy angular correlation function and power spectrum \\ in constraining 
primordial non-Gaussianity}

\author{Ryo~Terasawa\orcidlink{0000-0002-1193-623X}}
\email{ryo.terasawa@ipmu.jp}
\affiliation{%
Kavli Institute for the Physics and Mathematics of the Universe (WPI),
 The University of Tokyo Institutes for Advanced Study (UTIAS),
 The University of Tokyo, Chiba 277-8583, Japan
}%
\affiliation{%
Department of Physics, Graduate School of Science,
 The University of Tokyo, 7-3-1 Hongo, Bunkyo-ku, Tokyo 113-0033, Japan
}%
\affiliation{%
Center for Data-Driven Discovery (CD3), Kavli IPMU (WPI), UTIAS, The University of Tokyo, Kashiwa, Chiba 277-8583, Japan
}%

\author{Yue~Nan\orcidlink{0000-0003-4720-2307}}
\author{Masahiro~Takada\orcidlink{0000-0002-5578-6472}}
\affiliation{%
Kavli Institute for the Physics and Mathematics of the Universe (WPI),
 The University of Tokyo Institutes for Advanced Study (UTIAS),
 The University of Tokyo, Chiba 277-8583, Japan
}%
\affiliation{%
Center for Data-Driven Discovery (CD3), Kavli IPMU (WPI), UTIAS, The University of Tokyo, Kashiwa, Chiba 277-8583, Japan
}%

\begin{abstract}
We investigate the angular power spectrum ($C_\ell)$ and angular correlation function ($w(\theta)$)
of galaxy number density field in the presence of the local-type primordial non-Gaussianity (PNG), 
explicitly accounting for the {\it integral constraint} in an all-sky survey.
We show that the PNG signature in $C_{\ell}$ is confined to low multipoles in the linear regime, whereas 
its signature in $w(\theta)$ extends across a wide range of angular scales, including 
those below the nonlinear scale. 
Therefore, the equivalence between $C_\ell$ and $w(\theta)$
can be violated when scale cuts of multipoles or angular scales -- for example, to mitigate systematic effects -- are applied in the analysis.
Assuming samples of photometric galaxies divided into multiple redshift bins
in the range $0<z<7$,
we forecast the precision of constraining the PNG parameter 
($f_{\rm NL}$) from the hypothetical measurements of  $C_\ell$ or $w(\theta)$ assuming
different scale cuts in the multipoles or angular scales, respectively. 
Our results imply that the PNG information can be extracted from $w(\theta)$ on relatively small angular scales
such as $\lesssim 10$~degree for a high-redshift galaxy sample  or from $w(\theta)$ measured in a survey with partial area coverage. 
\end{abstract}

\maketitle

\section{Introduction}
\label{sec:introduction}

The standard cosmological model, known as the $\Lambda$CDM model, assumes adiabatic and Gaussian primordial fluctuations as the initial seeds for all cosmic structures observed today \citep[e.g.,][]{Dodelson2nd}.
Simple inflation models {\em generally}
provide a compelling mechanism for generating such Gaussian primordial fluctuations \citep{1982PhRvL..49.1110G,2003JHEP...05..013M}.
Therefore, primordial non-Gaussianity (PNG) is key to understanding the physics of inflation and provides crucial insights for distinguishing between competing inflation models \citep{bartolo04}.

Observations of cosmic microwave background (CMB) anisotropies and large-scale structures serve as powerful probes
of the statistical properties of primordial fluctuations
\citep{bartolo04,dalal08,2014arXiv1412.4872D}. 
The simplest PNG model that can be tested with data 
is the {\em local-type} PNG characterized by the $f_{\rm NL}$ parameter, 
defined as
\begin{align}
    \Phi(\bm x) = \phi(\bm x) + f_{\rm NL} \left(\phi^2 (\bm x) - \langle \phi^2\rangle \right),
\end{align}
where $\Phi(\bm x)$ is the PNG perturbation field (such as the primordial curvature perturbation), $\phi(\bm x)$ is the Gaussian perturbation field, and 
the strength of PNG is parametrized by $f_{\rm NL}$.
{\em Any} detection of $f_{\rm NL}$ will {\em rule out} single-field inflation \citep{2004JCAP...10..006C,2013PhRvD..88h3502P}. Therefore, pursuing 
the most stringent constraints on  $f_{\rm NL}$ 
from data is crucial for testing inflation models.

The seminal work by Dalal et al.~\cite{dalal08} opens up a window for searching for the PNG signature using 
large-scale structure data instead of CMB data \citep[also see][]{2009PhRvD..79l3507W,Desjacques18}. They demonstrated that the local-type PNG induces a {\em characteristic} scale-dependent bias for large-scale structure tracers such as galaxies, which can be probed using the lowest-order clustering statistics, i.e., two-point statistics, without relying on higher-order moments.

Following this pioneering work, various attempts have been made to search for the PNG signature from wide-area galaxy surveys \citep{slosar08,2012ApJ...761...14H,2013MNRAS.428.1116R,2021MNRAS.506.3439R, 2023arXiv230202925K, 2023arXiv230915814C}. Most recently, Refs.~\cite{2023MNRAS.523..603R, 2023arXiv230701753R} used  angular clustering measurements of the photometric luminous red galaxy (LRG) sample, constructed from 
the Dark Energy Spectroscopic Instrument (DESI)
imaging surveys 
over 14,000~deg$^2$ in the redshift range $0.2<z<1.35$, 
 to constrain $f_{\rm NL}$.
Ref.~\cite{2024arXiv241117623C} used 3D clustering measurements from the spectroscopic samples of 
LRGs and quasars over the combined redshift range $0.6<z<3.1$ from the DESI Year~1 data to obtain
$f_{\rm NL} = {-3.6}_{-9.1}^{+9.0}$at $68\%$ confidence.  

In this paper, we address a rather basic question in the pursuit of the PNG signature from wide-area galaxy survey data:
Are configuration-space and Fourier-space analyses in the search for PNG equivalent \citep[also see Ref.][for the similar discussion]{slosar08}? To do this, we consider the angular power spectrum (therefore harmonic space corresponding to Fourier space) and the angular correlation function for a galaxy sample in a {\em full-sky} survey, as we can derive mathematically exact equations for the relevant clustering quantities.
We will pay attention to employing the {\em integral constraint} in deriving the equations as is done in actual measurements. We will show that the angular power spectrum and the angular correlation function are indeed equivalent to each other, {\em only if} we can use the information on all available angular scales.
However, the equivalence is not obvious in practice, as we often adopt ``scale cuts'' in the analysis, e.g. to avoid scales affected by poorly understood nonlinear effects and to mitigate systematic effects. 
We will investigate which scales in the angular power spectrum and the angular correlation function carry information about PNG, and forecast the precision of constraining $f_{\rm NL}$ for each of the counterpart quantities, assuming photometric galaxy samples are divided into multiple redshift bins when applying different scale cuts in each space.

We organize this paper as follows. In Section~\ref{sec:preliminary}, we introduce the model for the galaxy angular power spectrum 
in the presence of PNG.
In Section~\ref{sec:integral_constraint}, we 
introduce the angular correlation function and
note the integral constraint which is important when considering the local-type PNG.
In Section~\ref{sec:fisher}, we explain the details of the fisher analysis presented in this paper, and we will show the Fisher forecasts for the $f_{\rm NL}$ estimation in Section~\ref{sec:fisher_result}.
Section~\ref{sec:conclusion} is devoted to 
discussion and conclusion.

\section{Angular power spectrum with PNG}
\label{sec:preliminary}

In this paper, we consider the 2D number density fluctuation field of galaxies on the sphere 
in the $i$-th tomographic redshift bin, 
denoted as $\Delta^i_{\rm g}$:
\begin{align}
\Delta^i_{\rm g}(\bhat{n})=\int_0^{\chi_H}\!\mathrm{d}\chi~W^i(\chi)\delta_{\rm g}(\chi,\chi\bhat{n}),
\end{align}
where we assumed that an observer is at the coordinate origin, 
$\bhat{n}$ is the unit vector on the sphere, $\chi$ is the comoving radial distance to redshift $z$ via the distance-redshift relation $\chi=\chi(z)$, and
$\chi_H$ is the distance to the Hubble horizon. 
Throughout this paper, we assume a flat-geometry $\Lambda$CDM cosmology.
Here we consider tomographic redshift bins of galaxies where galaxies are divided into subsamples in redshift bins, e.g. based on photometric redshift information, and $W^i$ denotes the selection function in the $i$-th redshift bin, 
defined to satisfy the normalization $\int^{\chi_H}_0\!\mathrm{d}\chi~W^i(\chi)=1$.

We employ the following model of the 3D power spectrum for galaxies at two radial distances, 
$\chi_1$ and $\chi_2$ (or equivalently at two redshifts $z_1$ and $z_2$), 
in the presence of PNG: 
\begin{align}
P_{\rm gg}(k;\chi_1,\chi_2)&\equiv 
\left[b_1(\chi_1)+\Delta b(k;\chi_1,f_{\rm NL})\right]\nonumber\\
&\hspace{2em}\times\left[b_1(\chi_2)+\Delta b(k;\chi_2,f_{\rm NL})\right]\nonumber\\
&\hspace{2em}\times
D_0(\chi_1)D_0(\chi_2)P^L_{\rm mm}(k;z=0),
\label{eq:pk_def}
\end{align}
where 
$b_1(\chi)$ is the linear bias parameter of galaxies in a sample
at 
redshift $z(\chi)$,
$\Delta b(k)$ is the scale-dependent bias function induced by the PNG effect (see below), 
and $P^L_{\rm mm}(k;z=0)$ is the linear matter power spectrum today.
We denote the linear growth factor that is normalized to the scale factor during the matter-dominated era as $D(z)$, and 
$D_0(z) \equiv D(z)/D(z=0)$ is the 
linear growth factor 
normalized today
as $D_0(z=0) = 1$.
Note that 
the amplitude of the linear matter power spectrum is normalized by the 
amplitude today, i.e., using the $\sigma_8$ normalization.  
Throughout this paper, we assume the linear theory; or in other words, we consider the scales where the linear theory is valid. 
When PNG is present, the galaxy bias becomes scale-dependent, rather than constant, 
on large scales \citep{dalal08}:
\begin{align}
 \Delta b(k; \chi,f_{\rm NL}) &= 2(b_1-p)f_{\rm NL} \frac{\delta_c}{\alpha(k)},
 \label{eq:b1_fnl_def}
\end{align}
where $\delta_c=1.686$, $p$ is the constant for which we employ $p=1$ (see Ref.~\cite{2019Galax...7...71B} for a review), and $\alpha(k)$ given as 
\begin{align}
 {\alpha(k)} = \frac{2k^2 T(k) D(z)} {3\Omega_{\rm m} H_0^2} 
 \label{eq:alpha_def}
\end{align}
relates the matter density field $\delta_{\rm m}$ and the potential $\Phi$ through $\delta_{\rm m}(k) = \alpha(k) \Phi(k)$. 
The Hubble constant $H_0= 100h~{\rm Mpc}^{-1}{\rm km}~{\rm s}^{-1}$, where $h$ is
 the Hubble parameter,
and $\Omega_{\rm m}$ is the density parameter of matter.
$T(k)$ is the transfer function of matter fluctuations at $z=0$, which has an asymptotic large-scale limit 
of $T(k)\rightarrow 1$ at $k\rightarrow 0$. $\Delta b(k)$ has an infrared-divergent behavior of $
\Delta b\propto k^{-2}$ at 
the limit $k\rightarrow 0$.

The angular power spectrum between galaxies in the $i$-th and $j$-th tomographic bins
is given as
\begin{align}
    C^{ij}_{{\rm gg},\ell} &= 
    4\pi \int_0^{\chi_H}\!\! \mathrm{d}\chi_1W^i(\chi_1)\int_0^{\chi_H}\!\!\mathrm{d}\chi_2
    W^j(\chi_2) \nonumber\\
    &\times \int\!\! \frac{k^2\mathrm{d}k}{2\pi^2} P_{\rm gg}(k; \chi_1,\chi_2)j_{\ell}(k\chi_1) j_{\ell}(k\chi_2) , \label{eq:cell}
\end{align}
where 
$j_\ell(x)$ is the spherical Bessel function of the order $\ell$.
As can be found from Eqs.~(\ref{eq:pk_def}) and (\ref{eq:b1_fnl_def}), the functions of time~($\chi$) and scale~($k$)
in each of the different terms of $P_{\rm gg}$ are multiplicative. 
{We can use the {\tt FFTlog}~\citep{2020JCAP...05..010F} to perform a computation of the $k$-integrals in Eq.~(\ref{eq:cell}).}

In the presence of the PNG effect ($f_{\rm NL}\ne 0$), the monopole moment of the angular power spectrum has an apparent divergence as 
\begin{align}
C_{{\rm gg},\ell=0}\sim k^3[\Delta b(k)]^2 P^L_{\rm mm}(k)|_{k\rightarrow 0}
\sim k^{n_s-1}|_{k\rightarrow 0},
\label{eq:apparent_divergence}
\end{align}
where we have use the facts that $j_{\ell=0}\rightarrow 1$ and $P_{\rm mm}^{L}\sim 
k^{n_s}$
at $k\rightarrow 0$
and that the $\chi_1$- (or $\chi_2$-) integral gives a constant multiplicative factor (no $k$-dependence). 
Since the spectral index of the primordial curvature power spectrum,  $n_s\simeq 0.96$, $C_{{\rm gg},\ell=0}$
has an infrared divergence. 
The power spectrum at multipoles other than $\ell=0$ does not have any divergence. 
This apparent divergence can be removed by imposing the integral constraint as we discuss below.

Before going to the next section, we would like to note that the galaxy-matter power spectrum, denoted as $P_{\rm gm}$, does not have such a strong scale-dependent 
modification for the limit of $k\rightarrow 0$, in contrary to $P_{\rm gg}$. The cross-power spectrum is observable, e.g. 
via galaxy-galaxy weak lensing \citep[e.g.][]{2022PhRvD.105b3520A, 2023PhRvD.108l3517M}. 
Since $P_{\rm gm}\propto \Delta b(k)$ at the limit of $k\rightarrow 0$, we can compute the asymptotic behavior of $P_{\rm gm}$,
similarly 
 to Eq.~(\ref{eq:apparent_divergence}):
\begin{align}
\left.k^3 P_{\rm gm}(k)\right|_{k\rightarrow 0} \sim \left.k^{n_s+1}\right|_{k\rightarrow 0}\rightarrow 0. 
\end{align}
Thus, $k^3P_{\rm gm}$ does not have a divergence at $k\rightarrow 0$.
Note that $k^3P_{\rm gm}$ is related to the two-point correlation function of galaxy-galaxy weak lensing. 
We would also like to recall that the nonlinear matter power spectrum $P_{\rm mm}$, which is observable via 
cosmic shear \citep[e.g.,][]{Terasawa24},  
is not affected by the PNG.
Summarizing these dependencies, we can find that, in the presence of the PNG effect,  
the so-called $2\times$2pt analysis combining galaxy-galaxy weak lensing 
and galaxy clustering, equivalently the information in $P_{\rm gm}$ and $P_{\rm gg}$ cannot generally constrain 
cosmological parameters in an unbiased manner by resolving the galaxy bias uncertainty simultaneously, because 
the PNG effect violates the simple relations, $P_{\rm gm}\propto b_1 P_{\rm mm}$ and $P_{\rm gg}\propto b_1^2 P_{\rm mm}$, in the linear regime. 
We should keep in mind this possible physical limitation for future surveys.

\section{Integral constraint}
\label{sec:integral_constraint}

As discussed by Eq.~(\ref{eq:apparent_divergence}),
the angular power spectrum at $\ell=0$
has an apparent infrared divergence\footnote{Note that Ref.~\cite{2012PhRvD..85b3504J} discussed the general relativistic corrections to the galaxy power spectrum on very large scales that 
are close to the horizon scale, but we do not consider the effect throughout this paper.}.
However, 
the monopole power spectrum we can observe from actual data should not 
have such a divergence, because the number of galaxies we observe is always {\em finite}. 
The usual procedures we take for an actual measurement are as follows: i) define the {\em mean} angular number density of galaxies in a survey area, ii) calculate the number density fluctuation field relative to the mean number density, and then iii) measure the angular power spectrum. 
Thus we can define the ``observable'' angular power spectrum by imposing the {\em integral constraint}: 
\begin{align}
\tilde{C}^{ii}_{{\rm gg},\ell}\equiv C^{ii}_{\rm gg,\ell}-C^{ii}_{{\rm gg},\ell=0}\delta^K_{\ell 0},
\label{eq:tildeCell_def}
\end{align}
where $\delta^K_{\ell\ell'}$ is the Kronecker delta function: $\delta^K_{\ell\ell'}=1$ if $\ell=\ell'$, otherwise zero. The above power spectrum has a vanishing monopole by definition, i.e.
$\tilde{C}^{ii}_{{\rm gg},\ell=0}=0$, in contrary to Eq.~(\ref{eq:apparent_divergence}).
{Thus, the integral constraint corresponds to the fact that we give up observing the monopole from the data because the mean number density, 
i.e. the monopole component in the galaxy distribution needs to be estimated from the data itself, which is generally different from the true mean, 
as we discussed above.
} In the following, 
we consider only the auto-power spectrum or auto-correlation function of galaxies in the same 
redshift bin, i.e. $i=j$, for simplicity.

The angular correlation function is formally defined as
\begin{align}
w^{ii}_{\rm gg}(\theta)=\sum_\ell\frac{2\ell+1}{4\pi}C_{{\rm gg},\ell}^{ii} P_\ell(\cos\theta),
\label{eq:Cell_to_wtheta_def}
\end{align}
where $P_\ell(\mu)$ is the Legendre polynomials. 
The integral constraint for the angular correlation function for an all-sky survey
is given as 
\begin{align}
 \frac{1}{2}\int^{1}_{-1}\mathrm{d}(\cos\theta)~ \tilde{w}^{ii}_{\rm gg}(\theta)=0,
\label{eq:wtheta_integral_constraint}
\end{align}
where we use the tilde top symbol ``${~}\widetilde{~} {~}$'' to denote the angular correlation function $\tilde{w}^{ii}_{\rm gg}(\theta)$ on which 
the integral constraint is imposed. 
It is straightforward to express the ``observed'' correlation function, $\tilde{w}_{\rm gg}^{ii}(\theta)$, 
in terms of the original function, $w_{\rm gg}^{ii}(\theta)$ in Eq.~(\ref{eq:Cell_to_wtheta_def}), as
\begin{align}
\tilde{w}^{ii}_{\rm gg}(\theta)&=w^{ii}_{\rm gg}(\theta)-\frac{1}{2}\int_{-1}^1\!\mathrm{d}(\cos\theta)~ 
w^{ii}_{\rm gg}(\theta)\nonumber\\
&= w^{ii}_{\rm gg}(\theta)-\frac{1}{4\pi}C^{ii}_{{\rm gg},\ell=0}\nonumber\\
&= \sum_{\ell\ge 1} \frac{2\ell+1}{4\pi}C^{ii}_{{\rm gg},\ell}P_\ell(\cos\theta).
\label{eq:wobs_def}
\end{align}
One can easily check that
the integral 
constraint Eq.~(\ref{eq:wtheta_integral_constraint}) is satisfied. 
In the second equality on the r.h.s. we inserted Eq.~(\ref{eq:Cell_to_wtheta_def}) and then 
used the orthogonal relation of the Legendre polynomials, given as $\int_{-1}^1\mathrm{d}\mu/2~P_\ell(\mu)P_{\ell'}(\mu)=\delta^K_{\ell\ell'}/(2\ell+1)$.
The PNG affects $\tilde{w}_{\rm gg}^{ii}(\theta)$ through 
the first and second terms in the second equality, or through $C_{{\rm gg},\ell}^{ii}$ at $\ell\ge 1$ as can be found from Eq.~(\ref{eq:wobs_def}).

Comparing the third line on the r.h.s. and Eq.~(\ref{eq:tildeCell_def}), we can find the following relation: 
\begin{align}
\tilde{w}^{ii}_{\rm gg}(\theta)=\sum_{\ell\ge 0}\frac{2\ell+1}{4\pi}\tilde{C}_{{\rm gg},\ell}P_\ell(\cos\theta).
\label{eq:tildew_tildeCell_equivalence}
\end{align}
As expected, this equation shows that the ``observed'' correlation function $\tilde{w}_{\rm gg}^{ii}(\theta)$
and the ``observed'' angular power spectrum $\tilde{C}_{{\rm gg},\ell}^{ii}$ are equivalent.

There is a notable difference between the integral constraint in the power spectrum and the correlation function, 
or harmonic space and real space. As can be found from Eq.~(\ref{eq:tildeCell_def}), 
the integral constraint for the power spectrum appears only at its monopole 
moment ($\ell=0$). In this sense, the integral constraint in harmonic space is {\em local}. 
On the other hand, the second term in the first line on the r.h.s. of Eq.~(\ref{eq:wobs_def}) is the integral 
constraint and gives an additive correction to the underlying true correlation function. 
Thus, the integral constraint in real space is {\em non-local} in the sense that the correction induces 
a scale-dependent change in 
the correlation function over a range of separation scales ($\theta$). 
In addition, in actual observations, we usually employ scale cuts or have access to a limited range of 
multipoles or separation scales, e.g. due to a partial sky coverage or to mitigate the systematic effects. 
Therefore, whether the harmonic-space and real-space analyses 
are equivalent is not obvious. We address this question in this paper.

\section{Fisher analysis}
\label{sec:fisher}

We consider a realistic setup of galaxy surveys to 
study how the PNG effect can be measured using different scale cuts
in harmonic-space and real-space analyses. 
In this section, we describe setups used to perform the Fisher forecasts. 
{Throughout the paper, we assume the fiducial cosmology to be a flat $\Lambda$CDM model determined by $\{A_s, n_s, \omega_{\rm b}, \omega_{\rm c}, h\} = \{2.19 \times 10^{-9}, 0.9645, 0.02226, 0.12055, 0.6727\}$, which is consistent with the {\it Planck} 2015 best-fit cosmology~\citep{planck-collaboration:2015fj}. 
$A_s$ and $n_s$ are the amplitude and the spectral index parameters of the primordial curvature power spectrum at 
pivot scale $k_{\rm pivot}=0.05~{\rm Mpc}^{-1}$, $h$ is the Hubble constant parameter, and $\omega_{\rm b}(\equiv \Omega_{\rm b}h^2)$ is the physical density parameter of baryons. 
The physical density parameter of CDM is given as $\omega_{\rm c}(\equiv \Omega_{\rm c}h^2)=\Omega_{\rm m}h^2-\omega_b-\omega_\nu$,
where $\omega_\nu$ is the physical density parameter of massive neutrinos.}
Note that the density parameters we choose satisfy the identity $\Omega_{\rm c}+\Omega_{\rm b}+\Omega_{\nu}=\Omega_{\rm m}$.

\subsection{Methodology}
\subsubsection{Galaxy redshift distribution}
\label{sec:dndz}

We consider an LSST~Y1-like survey following the DESC science requirements document (DESC SRD)~\cite{thelsstdarkenergysciencecollaboration2021lsst}. 
We consider two different setups for the redshift range similar to the Ref.~\cite{2018PhRvD..97l3540S}. First, 
we adopt the following redshift distribution for the {galaxies within redshift range $z = 0-4$}: 
\begin{align}
    \frac{dN}{dz} \propto z^2 \exp [-(z/0.26)^{0.94}],
    \label{eq:dndz_lowz}
\end{align}
normalized to satisfy
\begin{align}
    n_{\rm eff} = \int_0^{4.0} dz \frac{dN}{dz} = 18~{\rm arcmin}^{-2}.
\end{align}
This redshift distribution has $\langle z\rangle\simeq 0.9$ for the mean redshift. 
We assume that photometric redshifts of galaxies are used to define this galaxy sample.

For the higher redshift range $z = 4-7$, we consider a galaxy sample mimicking Lyman break galaxies (LBGs). 
This sample is motivated by the Subaru Hyper Suprime-Cam survey~\cite{GOLDRUSH_IV} that found
[1836244, 139359, 2567, 292]
LBG candidates using [$g,r,i,z$] dropout techniques, respectively, from the region with an area of
$307.9~{\rm deg}^2$. 
We assume that the LSST~Y1-like survey can achieve a similar survey of LBGs over a much larger area footprint.  
To make a quantitative estimate, we assume the number densities given by 
$1.66~{\rm arcmin}^{-2}$ in 
$4 \leq z \leq 4.5$, 
$0.126~{\rm arcmin}^{-2}$ in 
$4.5 \leq z \leq 5.5$,
$2.23\times 10^{-3}~{\rm arcmin}^{-2}$ 
in 
$5.5 \leq z \leq 6.5$,
and 
$2.63\times 10^{-4}~{\rm arcmin}^{-2}$ 
in the bin of $6.5 \leq z \leq 7$, 
respectively.
For each redshift range, we assume the constant $\mathrm{d}N/\mathrm{d}z$, normalized to satisfy 
the above number density.

We divide the galaxies into six tomographic redshift bins, given as 
$z =[0,0.5], [0.5,1], [1,2], [2,3], [3,4]$, and $[4,7]$, respectively, using the redshift distribution of galaxies that we described above. 
In the following, we consider only the auto-power spectrum of galaxies in the same redshift bin, and 
do not consider
the cross power spectra of galaxies residing in the different neighboring redshift bins, for simplicity. 
Since the main purpose of this paper is to study the difference in exploring the PNG signatures from 
the harmonic-space and real-space analyses, our treatment does not change the main conclusion. 
We assume the linear galaxy bias parameter with redshift dependence given by $b_1(z) = 1 + z$. Note that we ignore the evolution within each redshift bin, i.e. we treat the bias in each bin as constant, evaluated at the middle point between redshift bin boundaries.

\subsubsection{Covariance}
\label{sec:covariance}

The covariance matrix describes the statistical errors of an observable, here the power spectrum or the correlation function for each galaxy sample. We take analysis setups that would be used in an actual analysis. 
When we search for a PNG signal from data, 
we would assume a fiducial $\Lambda$CDM model with no PNG, i.e. $f_{\rm NL}=0$, 
to model the covariance matrix, because the $\Lambda$CDM model is one of the most successful models
and we want to look for any deviation from the standard model as a hint of new physics.  
Then we will check whether the measurement of $\tilde{C}_\ell$
or $\tilde{w}(\theta)$ shows a significant deviation from the $\Lambda$CDM model
expectation, due to the non-zero PNG effect.

We employ the Gaussian covariance matrix between the power spectra of different multipoles, 
$\ell$ and $\ell'$,
in each redshift bin, 
for the fiducial $\Lambda$CDM model with $f_{\rm NL}=0$~\citep{2011MNRAS.414..329C}:
\begin{align}
    {\bf C}[\tilde{C}_{{\rm gg},\ell}^{ii},\tilde{C}_{{\rm gg},\ell'}^{ii}]
    %{\rm Cov}(\ell, \ell') 
    = \frac{2\delta_{\ell \ell'}^{K}}{f_{\rm sky}(2\ell + 1)} \left( \tilde{C}_{{\rm gg},\ell}^{ii} + \frac{1}{\bar{n}_{\rm g}^i} \right)^2,
    \label{eq:cov_cell}
\end{align}
and we will assume $f_{\rm sky} = 1$ in the following. 
The Kronecker delta function
$\delta^K_{\ell\ell'}$ ensures that 
the power spectra of different multipole bins are independent from each other. 
We assume $f_{\rm NL}=0$ to compute $\tilde{C}_{{\rm gg},\ell}^{ii}$ in the covariance matrix. 
Throughout this paper, we adopt $\Delta \ell=1$ for the multipole binning 
in the calculations shown below.

The Gaussian covariance matrix between the 
angular correlation functions in the separation bins $\theta_a$ and $\theta_b$
can be computed from Eq.~(\ref{eq:cov_cell}) as 
\begin{align}
    {\bf C}[\tilde{w}_{\rm gg}^{ii}(\theta_a),\tilde{w}^{ii}_{\rm gg}(\theta_b)]
    &= \sum_{\ell,\ell'} \frac{(2\ell + 1)(2\ell' + 1)}{(4\pi)^2} \nonumber\\
    &\hspace{-4em}\times{P}_{\ell}(\cos{\theta_a}){P}_{\ell'}(\cos{\theta_b}) 
    {\bf C}[\tilde{C}_{{\rm gg},\ell}^{ii},\tilde{C}_{{\rm gg},\ell'}^{ii}].
    \label{eq:ACF_Cov}
\end{align}
Note that the angular correlation functions 
of different separation bins in the sample variance regime
are highly correlated with each other, even for the Gaussian covariance. These covariances (Eqs.~\ref{eq:cov_cell} and \ref{eq:ACF_Cov}) account for the integral constraint. 
Finally, it is useful to consider the shot noise regime of the covariance, where $\tilde{C}_{{\rm gg},\ell}^{ii}\ll
1/\bar{n}_{\rm g}^i$.
The shot noise contamination is given as 
\begin{align}
{\bf C}[\tilde{w}_{\rm gg}^{ii}(\theta_a),\tilde{w}^{ii}_{\rm gg}(\theta_b)]_{\rm sn}
&= \frac{1}{4\pi^2(\bar{n}_{\rm g}^i)^2}\sum_{\ell }\frac{2\ell +1}{2}P_\ell(\mu_a)P_\ell(\mu_b)\nonumber \\
&= \frac{1}{4\pi^2(\bar{n}_{\rm g}^i)^2}\delta_D(\mu_a-\mu_b)\nonumber\\
&\simeq \frac{1}{4\pi^2(\bar{n}_{\rm g}^i)^2 \sin\theta\Delta\theta}\delta^K_{ab}\nonumber\\
&=\frac{\delta^K_{ab}}{N_{\rm pair}(\theta_a;\Delta\theta)},
\end{align}
where $\mu_{a,b}\equiv \cos\theta_{a,b}$. 
$N_{\rm pair}(\theta_a;\Delta\theta)$ is the number of pairs separated by $\theta_a$ in the bin width $\Delta\theta$ for an all-sky survey: 
$N_{\rm pair}=4\pi\bar{n}_{\rm g}^i\times 2\pi\sin\theta_a\Delta\theta\bar{n}_{\rm g}^i/2$ \citep{2003MNRAS.344..857T,2008A&A...477...43J}, where $4\pi \bar{n}_{\rm g}^i$ is the total number of galaxies 
in the $i$-th redshift bin for an all-sky survey and
the division by 2 is needed to avoid the double counting of pairs. 
We also have used the approximation 
$\delta_D(\cos\theta_a-\cos\theta_b)=\delta_D(\theta_a-\theta_b)/\sin\theta_a\simeq \delta^K_{ab}/[\sin\theta_a \Delta\theta]$
when the discrete binning for $\theta_a$ is considered. 
The Kronecker delta function $\delta^K_{ab}$ ensures that the shot noise covariance elements for 
different separation bins $\theta_a$ and $\theta_b$ are independent of each other. 
In the following, we use the interpolation 
of the tabulated power spectra 
in the 
multipole range $\ell=[1,5\times 10^4]$ 
to compute the summation over multipoles
in $w_{\rm gg}(\theta)$ and the covariance matrix
(Eqs.~\ref{eq:wobs_def} and \ref{eq:ACF_Cov}). 
We employ 20~logarithmically-spaced bins, spaced by $\Delta\log\theta=0.2$, 
in the range $\theta=[10^{-3},10]~$deg.,
while we employ 30~linearly-spaced bins, spaced by $\Delta\theta\simeq5.66$~deg., 
 in the range 
$\theta=[10,180]$~deg.

\subsubsection{Fisher matrix}
\label{subsec:fisher_matrix}

Using the covariance matrix, we can compute the Fisher matrix for a hypothetical measurement 
of $C_{{\rm gg},\ell}^{ii}$ or $w_{\rm gg}^{ii}$ in
the $i$-th redshift bin, to forecast the precision of parameter estimation:
\begin{align}
F^{i}_{\alpha\beta}\equiv 
\frac{\partial {\bf d^T}}{\partial p_\alpha}
{\bf C}^{-1} \frac{\partial {\bf d}}{\partial p_\beta}
\label{eq:fisher}
\end{align}
where ${\bf d}$ is the data vector that is either of the power spectrum or the correlation function, i.e.
$C_{{\rm gg},\ell}^{ii}$ or $w_{\rm gg}^{ii}$, 
${\bf C}^{-1}$ is the inverse of the covariance matrix, and 
$p_{\alpha}$ is a set of parameters including $f_{\rm NL}$.
For the parameters, we consider $f_{\rm NL}$ and the linear bias parameter 
$b_1(\bar{z}_i)$ for the galaxy sample in each redshift bin. 
Along with the constraints using a single redshift bin, we also report the constraints by combining all the redshift bins as $F^{\rm All}_{\alpha\beta}\equiv \sum_{i} F^{i}_{\alpha\beta}$, where we ignored the cross-covariance between different redshift bins. In this combined analysis, we consider 7 parameters in total. 

Since the PNG effect appears only at large length scales in the linear regime, we include only the information of 
$C_{{\rm gg},\ell}^{ii}$ or $w_{\rm gg}^{ii}$ at scales in the linear regime. 
To define the linear {scale} in each redshift bin, 
we follow Ref.~\citep{Senatore_2015} and use the variance of matter fluctuations, defined as
\begin{align}
    \epsilon(\bar{z}_i) = \int_0^{k_{\rm NL}(\bar{z}_i)} \frac{d^3 k}{(2\pi)^3} P_{\rm mm}^{L}(k;\bar{z}_i).
    \label{eq:def_linearscale}
\end{align}
We adopt the threshold value of $\epsilon = 0.3$ 
to define the nonlinear wavenumber, $k_{\rm NL}$,  
at the representative redshift of each redshift bin: $\bar{z}_i
= [0.25, 0.75, 1.5, 2.5, 3.5, 5.5]$, 
where 
the higher-order corrections to the linear power spectrum at $k<k_{\rm NL}$
are up to 
$\sim 30\%$ ~\citep{Senatore_2015}.
We determine the 
corresponding angular 
scale
as 
$\ell_{\rm NL}(\bar{z}_i) = \pi/\theta_{\rm NL}(\bar{z}_i) \equiv k_{\rm NL}(\bar{z}_{i}) \chi(\bar{z}_{i})$.
When doing the Fisher forecasts for $C_{{\rm gg},\ell}$ or $w_{\rm gg}(\theta)$
as a function of $\ell_{\rm min}$ or $\theta_{\rm max}$, 
we include the information of $C_{{\rm gg},\ell}^{ii}$ at $\ell_{\rm min}\le \ell\le \ell_{\rm max}$
or $w_{\rm gg}^{ii}(\theta)$ at $\theta_{\rm min}\le \theta\le \theta_{\rm max}$, 
 where we set 
$\ell_{\rm max}\equiv \ell_{\rm NL}(\bar{z}_i)$ or $\theta_{\rm min}\equiv \theta_{\rm NL}(\bar{z}_i)$ 
in each redshift bin, respectively.

\subsection{Signal-to-noise ratio and the sensitivity to $f_{\rm NL}$}
\label{sec:sn}

\begin{figure*}
    \includegraphics[width=\columnwidth]{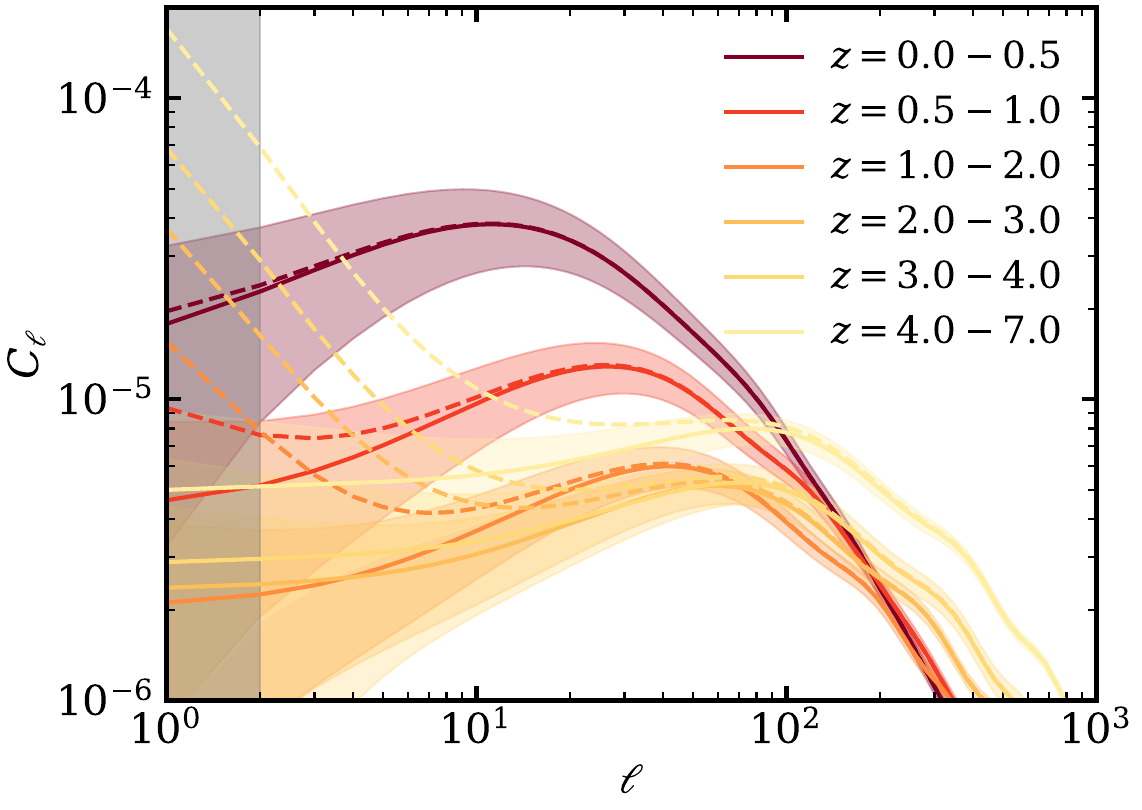}
    \includegraphics[width=\columnwidth]{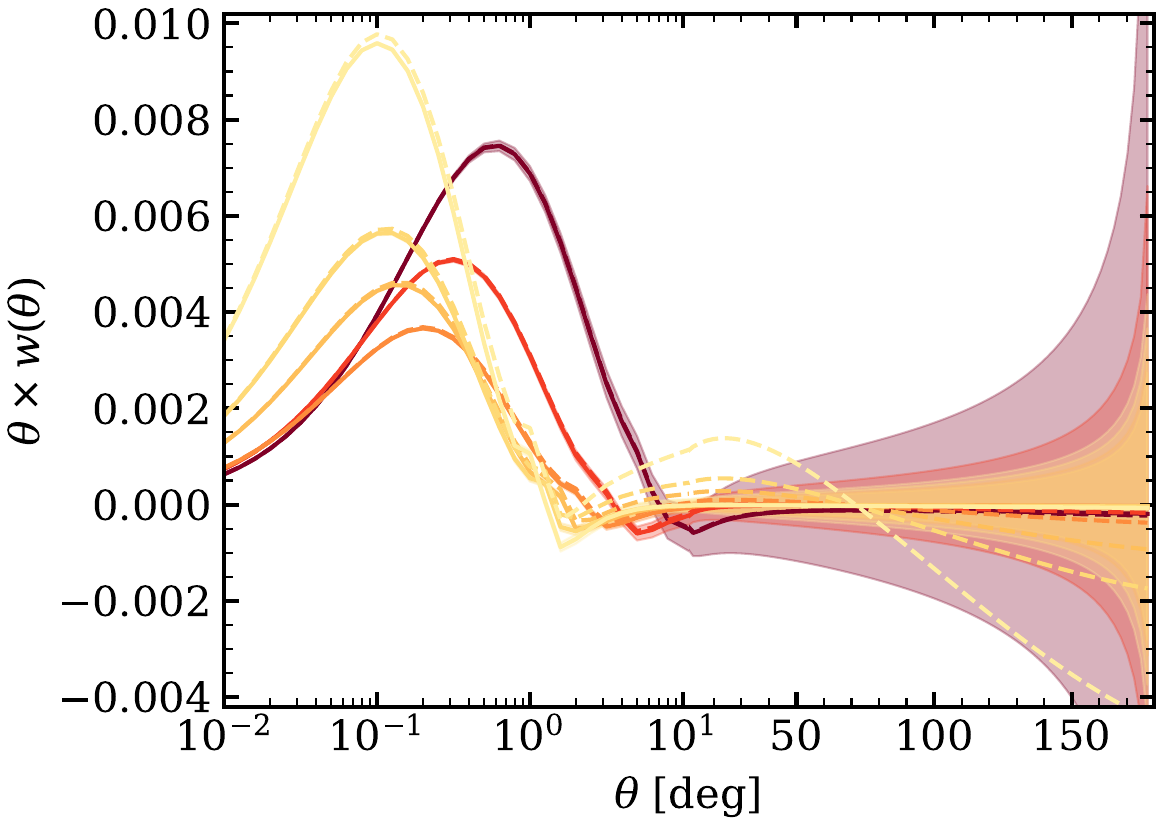}
    \caption{{\it Left panel}: Comparison of the angular power spectra for the hypothetical galaxy sample
        in each redshift bin (see Section~\ref{sec:dndz}) for an all-sky survey.
        The solid and dashed lines for each galaxy sample show the results  with and without 
    the local-type primordial non-Gaussianity, $f_{\rm NL}=30$ and 0, respectively. Other cosmological parameters 
    are fixed to the values of the $\Lambda$CDM model.
    The shaded region around the 
    solid line in each $\ell$ bin shows the expected
    $1\sigma$-measurement error computed from the diagonal terms of the
    Gaussian covariance matrix
    (Eq.~\ref{eq:cov_cell}), accounting for
    the shot noise for each galaxy sample and assuming
    $\Delta \ell=1$ for the multipole bin width.
        {\it Right panel}: Similarly to the left panel, but for the angular correlation functions.
    Here we adopt 20~logarithmically-spaced bins, spaced by $\Delta\log\theta=0.2$, in the range 
    $\theta=[10^{-3},10]$~deg. and 30~linearly-spaced bins, spaced by $\Delta\theta\simeq 5.66$~deg., in the 
    range $\theta=[10,180]$~deg.
}
    \label{fig:Cell-wtheta}
\end{figure*}
\begin{figure*}
    \includegraphics[width=\columnwidth]{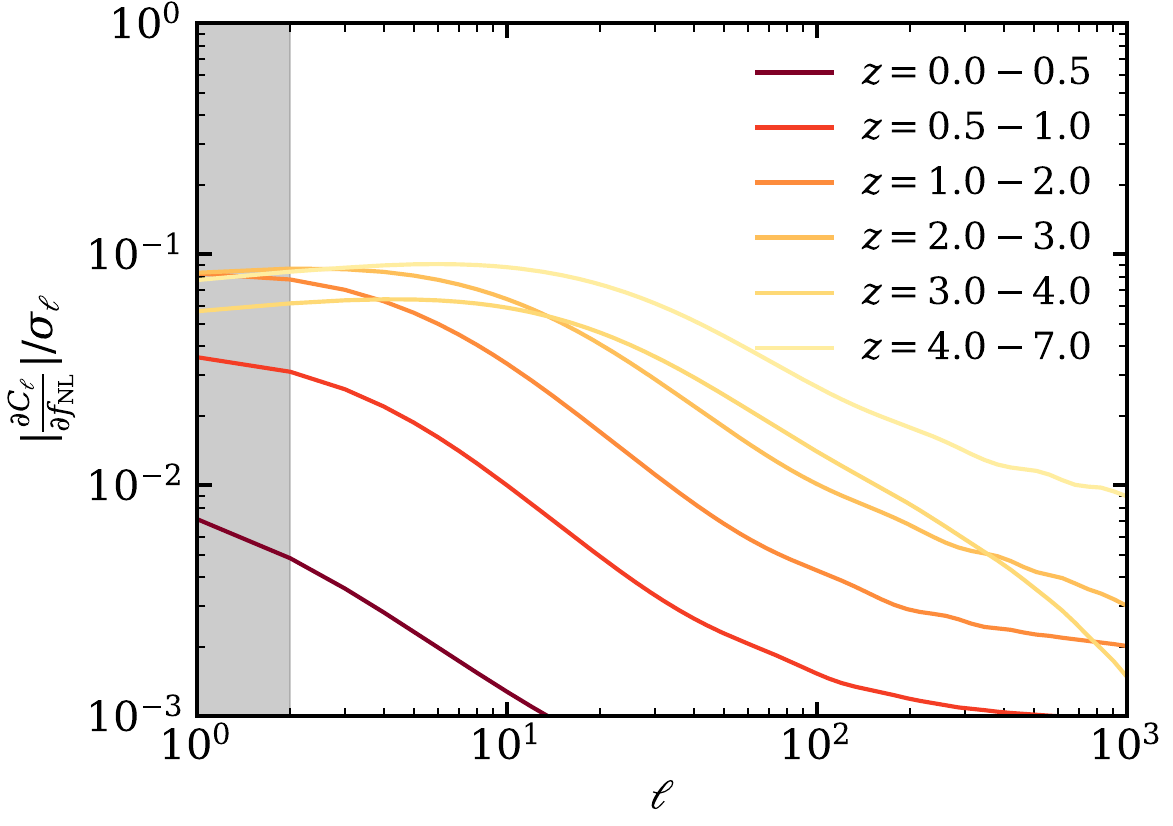}
    \includegraphics[width=\columnwidth]{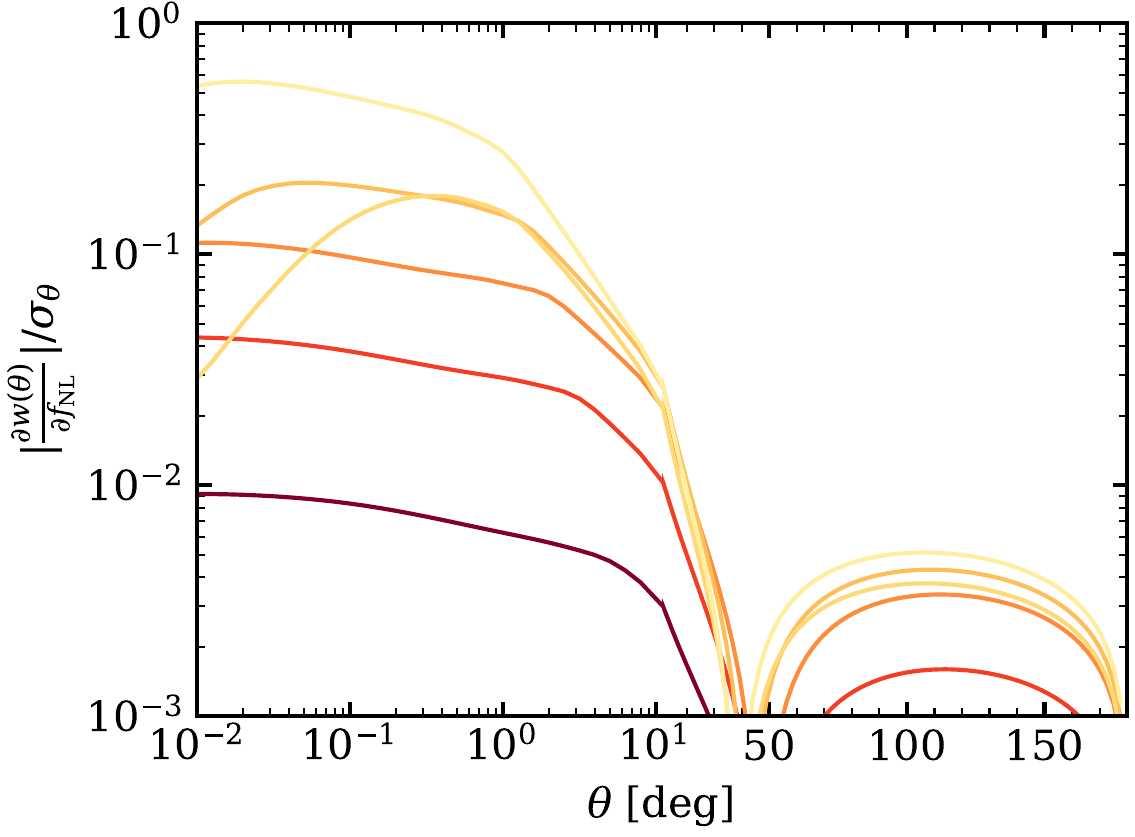}
    \caption{Sensitivity of the angular power spectra, $C^{ii}_{{\rm gg},\ell}$,
    and the angular correlation functions, $w^{ii}_{\rm gg}(\theta)$, to $f_{\rm NL}$ for the galaxy sample in each redshift bin
    in each multipole bin
    or each angular separation bin, respectively. To study the sensitivity, 
    we divide the fractional derivative of each observable with respect to $f_{\rm NL}$ by the 
   expected  $1\sigma$ measurement error, 
   where $\sigma$ is computed from the diagonal terms of the Gaussian covariance matrix, i.e. 
    $\sigma=\sqrt{{\rm diag}({\bf C})}$ in each multipole or angular separation bin. 
    Note that we used the same binning in Fig.~\ref{fig:Cell-wtheta}, and
    the error bars for $C_{{\rm gg},\ell}$ in different multipole bins are independent (uncorrelated), while the error bars between different bins of $w_{\rm gg}(\theta)$ are correlated with each other (only the small separation bins in the shot noise regime become uncorrelated). 
    Therefore, the left and right panels are not directly comparable, while the sensitivities at small- and large-scale bins within each panel can be compared. Due to the integral constraint, the monopole moment of 
    $C_{{\rm gg},\ell}$ is not observable, as indicated by the shaded region, and the angular correlation functions have a zero-crossing at $\theta\sim 50$~deg. in our setting. 
}
    \label{fig:Cell-wtheta-per-sigma}
\end{figure*}

In Fig.~\ref{fig:Cell-wtheta}, we show the angular power spectra and the angular correlation functions
for the hypothetical galaxy sample in each redshift bin, for the $\Lambda$CDM model with and without 
the PNG, $f_{\rm NL}=30$ and 0, respectively. For $C^{ii}_{{\rm gg},\ell}$ in 
harmonic
space, the PNG effect 
appears only at low multipoles
and becomes more significant
at lower multipoles. 
On the other hand, for $w_{{\rm gg}}^{ii}(\theta)$ in real space, the PNG effect appears over all the scales.
Comparing the results for the low and high redshift galaxy samples shows that the PNG effect is more significant at higher redshifts because a given angular scale at higher redshifts arises from the fluctuations of longer wavelengths, which are more affected by the PNG effect.

In Fig.~\ref{fig:Cell-wtheta-per-sigma}, we study $|\partial C^{ii}_{{\rm gg},\ell}/\partial f_{\rm NL}|/\sigma_\ell$ 
and $|\partial w^{ii}_{\rm gg}(\theta)/\partial f_{\rm NL}|/\sigma_\theta$ as an indicator of 
the sensitivity 
to $f_{\rm NL}$ as a function of multipole and angular-separation bins.
Here we used the diagonal terms of the covariance matrix for $C_{{\rm gg},\ell}$ 
and $w_{\rm gg}(\theta)$ to compute the $1\sigma$ error bar in each bin. 
Note that the error bars between the different bins for $w_{\rm gg}(\theta)$ are correlated, except for very small angular separation bins where the shot noise dominates. Therefore, we cannot directly compare the results in 
the left and right panels, and these figures are intended to compare the results at small and large multipole 
or angular separation bins in each panel. For $C_{{\rm gg},\ell}$, most of the sensitivity arises from 
the lower multipoles. Note that the monopole of $C_{{\rm gg},\ell}$ is not observable due to the integral constraint (Eq.~\ref{eq:tildeCell_def}). In addition, 
$C_{{\rm gg},\ell}$ at dipole and low multipoles around it display the higher signal-to-noise ratios, but the measurements require an all-sky or a wide-area survey.
On the other hand, $w_{\rm gg}(\theta)$ at small angular separations show a significant signal-to-noise, 
even on the nonlinear small scales, 
as a consequence of the integral constraint, which gives an additive correction to the underlying correlation function (see Eq.~\ref{eq:wobs_def}). In summary of Fig.~\ref{fig:Cell-wtheta-per-sigma}, the PNG 
affects $C_{{\rm gg},\ell}$ and $w_{\rm gg}(\theta)$ in different ways.

\subsection{Forecasts for $f_{\rm NL}$ estimation} 
\label{sec:fisher_result}

\begin{figure}
    \includegraphics[width=8cm]{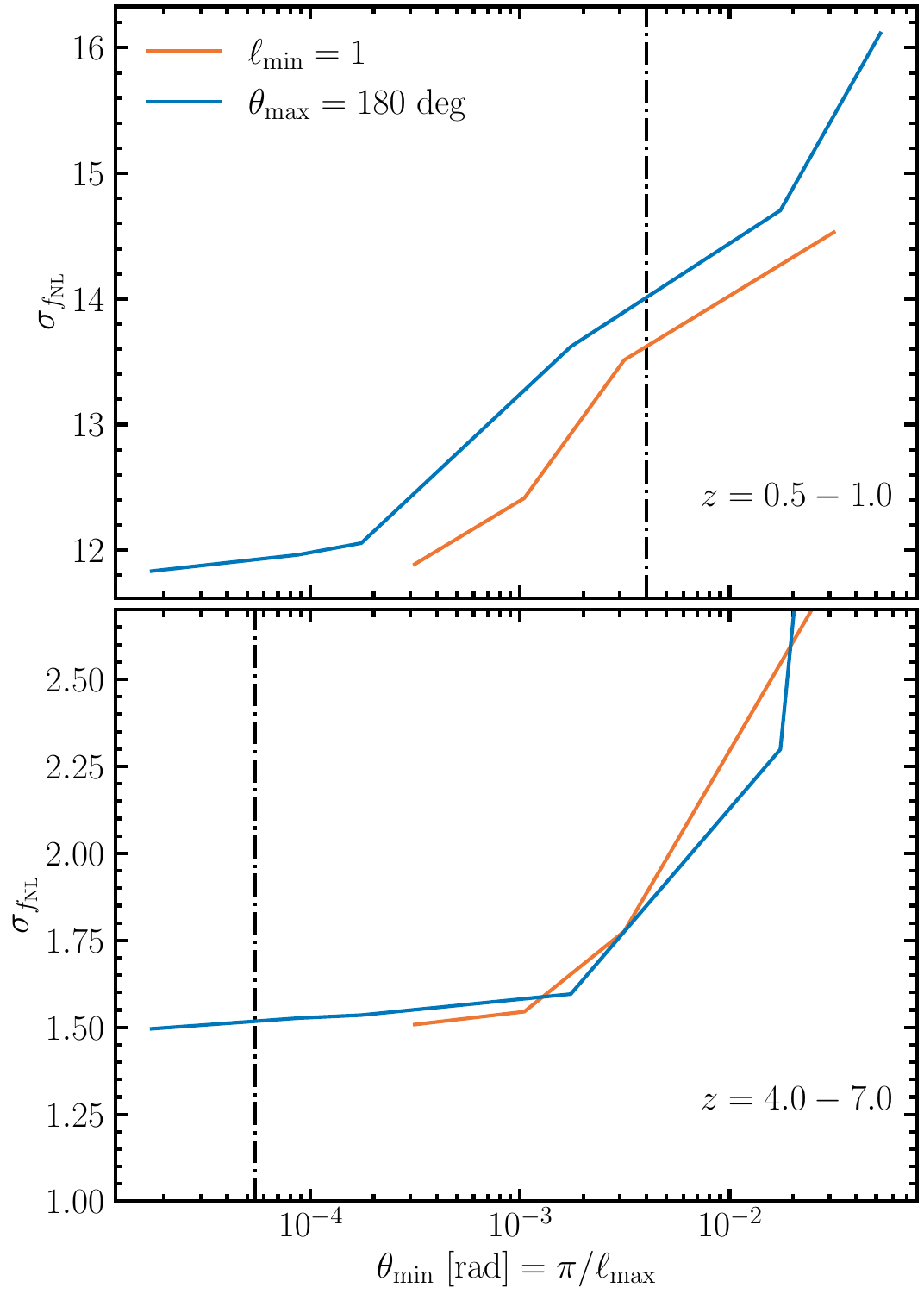}
\caption{The marginalized $1\sigma$ error of
$f_{\rm NL}$ computed using 
the Fisher analysis (see text); we include the information of $C_{{\rm gg},\ell}$ or $w_{\rm gg}(\theta)$ over 
the range of $\ell_{\rm min}\le\ell\le\ell_{\rm max}$ or $\theta_{\rm min}\le \theta\le \theta_{\rm max}$, as a function of the minimum 
angular scale $\theta_{\rm min}(=\pi/\ell_{\rm max})$, while keeping $\theta_{\rm max}=180$~deg. or $\ell_{\rm min}=1$ fixed.
Note that we assume an all-sky survey $f_{\rm sky}=1$, and 
the result for $\theta_{\rm min}\rightarrow0$, corresponding to $\ell_{\rm max}\rightarrow \infty$,
represents the case where all available angular scales are included.
The upper and lower panels show the results for the 
galaxy samples in the redshift ranges
$z=[0.5, 1.0]$ 
and $z=[4.0, 7.0]$, respectively.  
Vertical dash-dotted line in each panel
shows the angular scale corresponding to the nonlinear scale: $\ell_{\rm NL}(z) = \pi/\theta_{\rm NL}(z) = k_{\rm NL}(z) \chi(z)$
(see Eq.~\ref{eq:def_linearscale}). The nonlinear effects become significant on angular scales below this.
}
    \label{fig:sigma_thetamin}
\end{figure}
\begin{figure}
    \includegraphics[width=8cm]{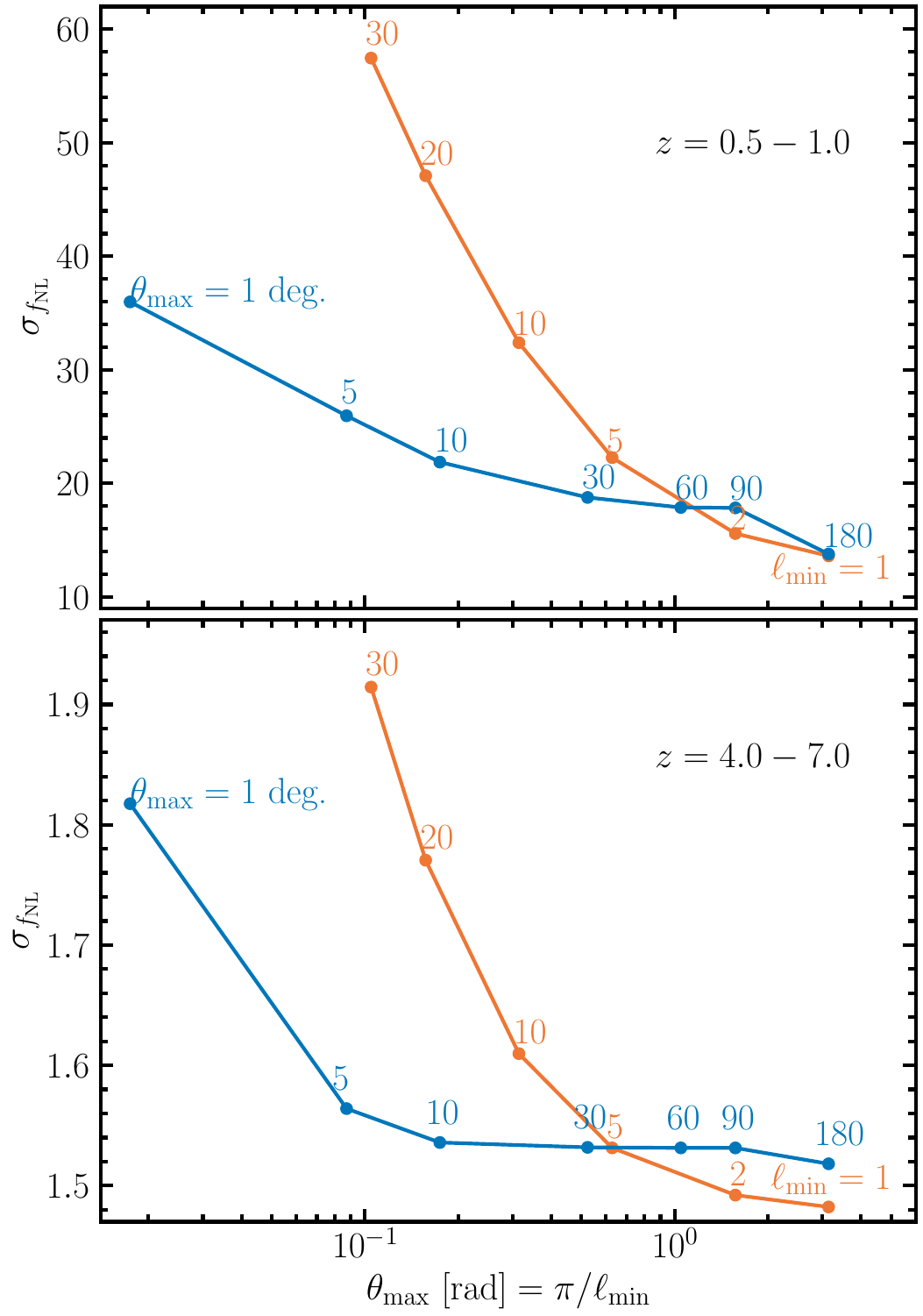}
\caption{Similar to the previous figure, but this figure shows the $1\sigma$ error of $f_{\rm NL}$ as a function of the maximum angular scale $\theta_{\rm max}(=
\pi/\ell_{\rm min})$. We keep the minimum angular scale fixed to the nonlinear scale, $\ell_{\rm max}=\ell_{\rm NL}$ (or $\theta_{\rm min}=\theta_{\rm NL}$) for each redshift slice; in other words, we included the information of $C_{{\rm gg},\ell}$ or $w_{\rm gg}(\theta)$ only on scales in the linear regime.
The number next to the dot symbol in each line denotes $\theta_{\rm max}$ or $\ell_{\rm min}$.
}
    \label{fig:sigma_thetamax}
\end{figure}

In this section, we show the main results of this paper.
As we demonstrated in Section~\ref{sec:integral_constraint}, $C_{{\rm gg},\ell}$ and $w_{\rm gg}(\theta)$ are equivalent, even in the presence of PNG. However, when scale cuts on multipoles or angular scales are applied, for instance, to avoid scales affected by 
poorly understood nonlinear effects
and/or mitigate systematic effect, the equivalence can be violated. In the following, we investigate  
the equivalence in constraining $f_{\rm NL}$ when including the information of $C_{{\rm gg},\ell}$ or 
$w_{\rm gg}(\theta)$ over a given range of angular scales, 
$l_{\rm min}\le \ell \le l_{\rm max}$ or $\theta_{\rm min}\le \theta\le \theta_{\rm max}$ in the Fisher analysis.
For the following discussion, we often use the relation $\theta=\pi/\ell$ to connect a given angular scale ($\theta$)
with the corresponding multipole ($\ell$), 
and vice versa.

Fig.~\ref{fig:sigma_thetamin} shows
the precision of constraining $f_{\rm NL}$ as a function of the minimum angular scales, 
$\theta_{\rm min}$ or $\ell_{\rm max}$, while keeping 
the maximum angular scale, $\theta_{\rm max}$ or $\ell_{\rm min}$, {\it fixed} in the Fisher forecast.
First, we consider the case of $\theta_{\rm max}=180$~deg. or $\ell_{\rm min}=1$, which corresponds to utilizing  all available scales on large scales. Note that $C_{{\rm gg},\ell}$ at the monopole $\ell=0$ is not observable due to the integral constraint. The figure shows that the marginalized errors
 of $f_{\rm NL}$ for 
$C_{{\rm gg},\ell}$ and $w_{\rm gg}(\theta)$ become equivalent to each other, 
{\em if} the $w_{\rm gg}(\theta)$ information down to small angular scales below the nonlinear scale (the vertical dot-dashed line) is included. By comparing the upper and 
lower panels corresponding to the results for the lower and higher redshift ranges, $z=[0.5,1.0]$ 
and $[4.0,7.0]$, respectively, we find that the galaxy sample in higher redshift bins provides tighter constraints on 
$f_{\rm NL}$ for a fixed $\theta_{\rm min}$ or $\ell_{\rm max}$, because a given angular scale corresponds to 
a larger length scale at higher redshifts compared to lower redshifts. The figure also shows that an all-sky survey can achieve the precision of $\sigma_{f_{\rm NL}}\sim 1$.
However, we note that the dipole moment of $C_{{\rm gg},\ell}$ at $\ell=1$ might be difficult to measure, because 
it is 
contaminated or even obscured by the Doppler or aberration effect caused by our own peculiar motion, which has an amplitude of $v/c\sim 10^{-3}$ \citep{1984MNRAS.206..377E,2002Natur.416..150B,2010PhRvD..82d3530I}.

In Fig.~\ref{fig:sigma_thetamax}, we investigate the precision of $f_{\rm NL}$ as a function of the 
maximum angular scale, $\theta_{\rm max}$ or $\ell_{\rm min}$, when the minimum angular scale, 
$\theta_{\rm min}$ or $\ell_{\rm max}$, is fixed to the nonlinear  scale (see around Eq.~\ref{eq:def_linearscale}), in the Fisher analysis.
Compared to Fig.~\ref{fig:sigma_thetamin}, $w_{\rm gg}(\theta)$ can capture 
the $f_{\rm NL}$ information more efficiently for the relatively small $\theta_{\rm max}$ compared to $C_{{\rm gg},\ell}$. This implies that 
the PNG signature can be explored from 
$w_{\rm gg}(\theta)$ at relatively small scales or $w_{\rm gg}(\theta)$ measured in a partial area survey.  It is interesting to find that 
the $w_{\rm gg}(\theta)$ information up to $\theta\simeq 10$ deg. 
for the high-redshift galaxy sample in $4<z<7$ can 
recover nearly all the information on $f_{\rm NL}$.
Note that, when including the information up to 
$l_{\rm min}=1$, $C_{{\rm gg},\ell}$ provides a tighter constraint on $f_{\rm NL}$ than $w_{\rm gg}(\theta)$, since we did not include 
the information of $w_{\rm gg}(\theta)$ on nonlinear scales by the setting.

\begin{table}
     \centering
     \begin{tabular}{c|c|c}
        \hline \hline
        redshift bin  &   $\sigma_{f_{\rm NL}}(\theta_{\rm max} = [10,180]~{\rm deg})$ &  $\sigma_{f_{\rm NL}}(\ell_{\rm min}= [1,30])$\\ \hline
      $[0.0, 0.5]$ & 170.40,~ 94.44 
            & 90.89,~ 546.99  \\ \hline
      $[0.5,1.0]$  & 21.87,~ 13.78 
                & 13.60,~ 57.46 \\ \hline
      $[1.0,2.0]$  & 6.63,~ 4.82 
                & 4.61,~ 12.34 \\ \hline
      $[2.0,3.0]$  & 3.13,~ 2.89 
                & 2.76,~ 5.31 \\ \hline
      $[3.0,4.0]$  & 2.96,~ 2.93 
            & 2.82,~ 4.57 \\ \hline
      $[4.0,7.0]$  & 1.54,~ 1.52 
            & 1.48,~ 1.91\\ \hline
      All  & 1.23,~ 1.18 
            & 1.14,~ 1.66 \\ \hline      
      \hline
     \end{tabular}
     \caption{
     The $1\sigma$ error of 
     $f_{\rm NL}$, 
     $\sigma_{f_{\rm NL}}$, for the galaxy samples in different redshift bin (see Section~\ref{sec:dndz}).  
    Here we consider the different maximum angular scale, $\theta_{\rm max}=[10,180]$~deg. for 
    $w_{\rm gg}(\theta)$, while $\ell_{\rm min}=[1,30]$ for $C_{{\rm gg},\ell}$. Note that we fixed the minimum angular scale to the nonlinear scale, similarly to Fig.~\ref{fig:sigma_thetamax}.
     The bottom row, labeled as ``All'', shows the result when all the galaxy samples are combined.
     }
     \label{tab:my_label}
 \end{table}
Finally, Table~\ref{tab:my_label} presents the marginalized $1\sigma$ error,
$\sigma_{f_{\rm NL}}$, as a function of the maximum scale considered, $\theta_{\rm max}$ or $\ell_{\rm min}$, for each redshift bin and all 6 redshift bins combined (denoted ``All" in the table).
The constraints are tighter for the higher redshift bins, reflecting the higher signal-to-noise at the higher redshift as we demonstrated in Section~\ref{sec:sn}.
Especially, the LBGs-like sample at $z=[4.0, 7.0]$ can reach $\sigma_{f_{\rm NL}} \sim 1.5$.
When combining all the redshift bins and using the smallest scale, we can obtain the desired accuracy of $\sigma_{f_{\rm NL}} \sim 1$.

\section{Conclusion}
\label{sec:conclusion}

In this paper, we have investigated the angular power spectrum ($C_{{\rm gg},\ell}$)
and the angular correlation function ($w_{\rm gg}(\theta)$) of the galaxy number density field in the presence of the local-type 
PNG, assuming an all-sky survey. We showed that 
the integral constraint, as done in actual measurements, is equivalent to the fact that we must refrain from measuring $C_{{\rm gg},\ell}$ at the monopole ($\ell=0$). In other words, this reflects the fact that we need to estimate the mean density of galaxies from data itself. 
We also demonstrated that the integral constraint eliminates the apparent infrared divergence in 
$C_{{\rm gg}, \ell=0}$
and
$w_{\rm gg}(\theta)$ caused by the PNG. 
Consequently, we showed that $C_{{\rm gg},\ell}$ and $w_{\rm gg}(\theta)$ are equivalent.

We showed that PNG affects $C_{{\rm gg},\ell}$ only at low multipoles, while its effect on $w_{\rm gg}(\theta)$ spans a wide range of angular scales including those below the nonlinear scale (see Figs.~\ref{fig:Cell-wtheta} and \ref{fig:Cell-wtheta-per-sigma}).
Therefore, we argued that the equivalence between $C_{{\rm gg},\ell}$ and $w_{\rm gg}(\theta)$ can be violated, if we adopt ``scale cuts'' as is often done in actual analyses to mitigate systematic effects. Interestingly, we showed that the PNG information can be probed from 
$w_{\rm gg}(\theta)$ at relatively small angular scales, or $w_{\rm gg}(\theta)$ measured from a partial sky-coverage survey. 
Since an all-sky galaxy survey is expensive or challenging due to foregrounds, such as contamination from the Milky Way, our results suggest that exploring PNG information through configuration-space statistics, rather than Fourier-space ones, could be a promising approach.
For example, with a galaxy sample at high redshifts, such as $4 < z < 7$ (e.g., Lyman break galaxies), we could constrain 
the PNG parameter $f_{\rm NL}$ at a precision of $\sigma_{f_{\rm NL}}\sim O(1)$, 
from measurements of $w_{\rm gg}(\theta)$ up to $\theta_{\rm max}\simeq 10$~degree, if we have a sufficiently wide area coverage. 
At least, our results suggest that exploring PNG through both Fourier-space and configuration-space clustering statistics is important.

Although we focus on angular clustering quantities in this paper, we believe that our results hold in general for three-dimensional clustering quantities measured from wide-area spectroscopic galaxy surveys such as the Subaru Prime Focus Spectrograph \cite{2014PASJ...66R...1T} and DESI. 
Since any survey is done in a finite volume, the mean density of galaxies has to be estimated from galaxies observed in the finite volume, and this imposes the integral constraint in the two-point clustering measurements. Then, the integral constraint affects the power spectrum only at the $k=0$ mode in the discrete Fourier transform as in Eq.~(\ref{eq:tildeCell_def}), while it provides an additive correction to the two-point correlation function, influencing the correlation function over a wide range of separation scales. 
Additionally, unlike angular clustering statistics, 3D clustering analysis allows us to use radial Fourier modes to probe PNG information.

Finally, we note that for a partial sky coverage or a finite volume spectroscopic survey, we have to account for the survey window effect
\citep[e.g.,][]{Hand_2018,2023arXiv230202925K}. This is rather straightforward, and we do not consider it challenging. Since it is extremely important to pursue local-type PNG at a precision of $\sigma_{f_{\rm NL}}<1$ (e.g., to test single-field inflation), we should 
explore all available methods. We hope that the results in this paper provide guidance for such directions.

\acknowledgments
We would like to thank Toshiki~Kurita and Sunao~Sugiyama for useful discussion. 
This work was supported in part by World Premier International Research Center Initiative (WPI Initiative), MEXT, Japan, JSPS KAKENHI Grant Numbers JP19H00677, 
JP20H05850, JP20H05855, JP20H05861,  JP23KJ0747, JP24K17041, and JP24H00215.

\bibliography{lssref}

\begin{thebibliography}{38}
\expandafter\ifx\csname natexlab\endcsname\relax\def\natexlab#1{#1}\fi
\expandafter\ifx\csname bibnamefont\endcsname\relax
  \def\bibnamefont#1{#1}\fi
\expandafter\ifx\csname bibfnamefont\endcsname\relax
  \def\bibfnamefont#1{#1}\fi
\expandafter\ifx\csname citenamefont\endcsname\relax
  \def\citenamefont#1{#1}\fi
\expandafter\ifx\csname url\endcsname\relax
  \def\url#1{\texttt{#1}}\fi
\expandafter\ifx\csname urlprefix\endcsname\relax\def\urlprefix{URL }\fi
\providecommand{\bibinfo}[2]{#2}
\providecommand{\eprint}[2][]{\url{#2}}

\bibitem[{\citenamefont{{Dodelson} and {Schmidt}}(2020)}]{Dodelson2nd}
\bibinfo{author}{\bibfnamefont{S.}~\bibnamefont{{Dodelson}}} \bibnamefont{and} \bibinfo{author}{\bibfnamefont{F.}~\bibnamefont{{Schmidt}}}, \emph{\bibinfo{title}{{Modern Cosmology}}} (\bibinfo{year}{2020}).

\bibitem[{\citenamefont{{Guth} and {Pi}}(1982)}]{1982PhRvL..49.1110G}
\bibinfo{author}{\bibfnamefont{A.~H.} \bibnamefont{{Guth}}} \bibnamefont{and} \bibinfo{author}{\bibfnamefont{S.~Y.} \bibnamefont{{Pi}}}, \bibinfo{journal}{\prl} \textbf{\bibinfo{volume}{49}}, \bibinfo{pages}{1110} (\bibinfo{year}{1982}).

\bibitem[{\citenamefont{{Maldacena}}(2003)}]{2003JHEP...05..013M}
\bibinfo{author}{\bibfnamefont{J.}~\bibnamefont{{Maldacena}}}, \bibinfo{journal}{Journal of High Energy Physics} \textbf{\bibinfo{volume}{2003}}, \bibinfo{eid}{013} (\bibinfo{year}{2003}), \eprint{astro-ph/0210603}.

\bibitem[{\citenamefont{{Bartolo} et~al.}(2004)\citenamefont{{Bartolo}, {Komatsu}, {Matarrese}, and {Riotto}}}]{bartolo04}
\bibinfo{author}{\bibfnamefont{N.}~\bibnamefont{{Bartolo}}}, \bibinfo{author}{\bibfnamefont{E.}~\bibnamefont{{Komatsu}}}, \bibinfo{author}{\bibfnamefont{S.}~\bibnamefont{{Matarrese}}}, \bibnamefont{and} \bibinfo{author}{\bibfnamefont{A.}~\bibnamefont{{Riotto}}}, \bibinfo{journal}{\physrep} \textbf{\bibinfo{volume}{402}}, \bibinfo{pages}{103} (\bibinfo{year}{2004}), \eprint{arXiv:astro-ph/0406398}.

\bibitem[{\citenamefont{{Dalal} et~al.}(2008)\citenamefont{{Dalal}, {Dor{\'e}}, {Huterer}, and {Shirokov}}}]{dalal08}
\bibinfo{author}{\bibfnamefont{N.}~\bibnamefont{{Dalal}}}, \bibinfo{author}{\bibfnamefont{O.}~\bibnamefont{{Dor{\'e}}}}, \bibinfo{author}{\bibfnamefont{D.}~\bibnamefont{{Huterer}}}, \bibnamefont{and} \bibinfo{author}{\bibfnamefont{A.}~\bibnamefont{{Shirokov}}}, \bibinfo{journal}{Physical Review D} \textbf{\bibinfo{volume}{77}}, \bibinfo{eid}{123514} (\bibinfo{year}{2008}), \eprint{0710.4560}.

\bibitem[{\citenamefont{{Dor{\'e}} et~al.}(2014)\citenamefont{{Dor{\'e}}, {Bock}, {Ashby}, {Capak}, {Cooray}, {de Putter}, {Eifler}, {Flagey}, {Gong}, {Habib} et~al.}}]{2014arXiv1412.4872D}
\bibinfo{author}{\bibfnamefont{O.}~\bibnamefont{{Dor{\'e}}}}, \bibinfo{author}{\bibfnamefont{J.}~\bibnamefont{{Bock}}}, \bibinfo{author}{\bibfnamefont{M.}~\bibnamefont{{Ashby}}}, \bibinfo{author}{\bibfnamefont{P.}~\bibnamefont{{Capak}}}, \bibinfo{author}{\bibfnamefont{A.}~\bibnamefont{{Cooray}}}, \bibinfo{author}{\bibfnamefont{R.}~\bibnamefont{{de Putter}}}, \bibinfo{author}{\bibfnamefont{T.}~\bibnamefont{{Eifler}}}, \bibinfo{author}{\bibfnamefont{N.}~\bibnamefont{{Flagey}}}, \bibinfo{author}{\bibfnamefont{Y.}~\bibnamefont{{Gong}}}, \bibinfo{author}{\bibfnamefont{S.}~\bibnamefont{{Habib}}}, \bibnamefont{et~al.}, \bibinfo{journal}{arXiv e-prints} \bibinfo{eid}{arXiv:1412.4872} (\bibinfo{year}{2014}), \eprint{1412.4872}.

\bibitem[{\citenamefont{{Creminelli} and {Zaldarriaga}}(2004)}]{2004JCAP...10..006C}
\bibinfo{author}{\bibfnamefont{P.}~\bibnamefont{{Creminelli}}} \bibnamefont{and} \bibinfo{author}{\bibfnamefont{M.}~\bibnamefont{{Zaldarriaga}}}, \bibinfo{journal}{\jcap} \textbf{\bibinfo{volume}{2004}}, \bibinfo{eid}{006} (\bibinfo{year}{2004}), \eprint{astro-ph/0407059}.

\bibitem[{\citenamefont{{Pajer} et~al.}(2013)\citenamefont{{Pajer}, {Schmidt}, and {Zaldarriaga}}}]{2013PhRvD..88h3502P}
\bibinfo{author}{\bibfnamefont{E.}~\bibnamefont{{Pajer}}}, \bibinfo{author}{\bibfnamefont{F.}~\bibnamefont{{Schmidt}}}, \bibnamefont{and} \bibinfo{author}{\bibfnamefont{M.}~\bibnamefont{{Zaldarriaga}}}, \bibinfo{journal}{\prd} \textbf{\bibinfo{volume}{88}}, \bibinfo{eid}{083502} (\bibinfo{year}{2013}), \eprint{1305.0824}.

\bibitem[{\citenamefont{{Wands} and {Slosar}}(2009)}]{2009PhRvD..79l3507W}
\bibinfo{author}{\bibfnamefont{D.}~\bibnamefont{{Wands}}} \bibnamefont{and} \bibinfo{author}{\bibfnamefont{A.}~\bibnamefont{{Slosar}}}, \bibinfo{journal}{\prd} \textbf{\bibinfo{volume}{79}}, \bibinfo{eid}{123507} (\bibinfo{year}{2009}), \eprint{0902.1084}.

\bibitem[{\citenamefont{{Desjacques} et~al.}(2018)\citenamefont{{Desjacques}, {Jeong}, and {Schmidt}}}]{Desjacques18}
\bibinfo{author}{\bibfnamefont{V.}~\bibnamefont{{Desjacques}}}, \bibinfo{author}{\bibfnamefont{D.}~\bibnamefont{{Jeong}}}, \bibnamefont{and} \bibinfo{author}{\bibfnamefont{F.}~\bibnamefont{{Schmidt}}}, \bibinfo{journal}{\physrep} \textbf{\bibinfo{volume}{733}}, \bibinfo{pages}{1} (\bibinfo{year}{2018}), \eprint{1611.09787}.

\bibitem[{\citenamefont{{Slosar} et~al.}(2008)\citenamefont{{Slosar}, {Hirata}, {Seljak}, {Ho}, and {Padmanabhan}}}]{slosar08}
\bibinfo{author}{\bibfnamefont{A.}~\bibnamefont{{Slosar}}}, \bibinfo{author}{\bibfnamefont{C.}~\bibnamefont{{Hirata}}}, \bibinfo{author}{\bibfnamefont{U.}~\bibnamefont{{Seljak}}}, \bibinfo{author}{\bibfnamefont{S.}~\bibnamefont{{Ho}}}, \bibnamefont{and} \bibinfo{author}{\bibfnamefont{N.}~\bibnamefont{{Padmanabhan}}}, \bibinfo{journal}{\jcap} \textbf{\bibinfo{volume}{8}}, \bibinfo{pages}{31} (\bibinfo{year}{2008}), \eprint{0805.3580}.

\bibitem[{\citenamefont{{Ho} et~al.}(2012)\citenamefont{{Ho}, {Cuesta}, {Seo}, {de Putter}, {Ross}, {White}, {Padmanabhan}, {Saito}, {Schlegel}, {Schlafly} et~al.}}]{2012ApJ...761...14H}
\bibinfo{author}{\bibfnamefont{S.}~\bibnamefont{{Ho}}}, \bibinfo{author}{\bibfnamefont{A.}~\bibnamefont{{Cuesta}}}, \bibinfo{author}{\bibfnamefont{H.-J.} \bibnamefont{{Seo}}}, \bibinfo{author}{\bibfnamefont{R.}~\bibnamefont{{de Putter}}}, \bibinfo{author}{\bibfnamefont{A.~J.} \bibnamefont{{Ross}}}, \bibinfo{author}{\bibfnamefont{M.}~\bibnamefont{{White}}}, \bibinfo{author}{\bibfnamefont{N.}~\bibnamefont{{Padmanabhan}}}, \bibinfo{author}{\bibfnamefont{S.}~\bibnamefont{{Saito}}}, \bibinfo{author}{\bibfnamefont{D.~J.} \bibnamefont{{Schlegel}}}, \bibinfo{author}{\bibfnamefont{E.}~\bibnamefont{{Schlafly}}}, \bibnamefont{et~al.}, \bibinfo{journal}{\apj} \textbf{\bibinfo{volume}{761}}, \bibinfo{eid}{14} (\bibinfo{year}{2012}), \eprint{1201.2137}.

\bibitem[{\citenamefont{{Ross} et~al.}(2013)\citenamefont{{Ross}, {Percival}, {Carnero}, {Zhao}, {Manera}, {Raccanelli}, {Aubourg}, {Bizyaev}, {Brewington}, {Brinkmann} et~al.}}]{2013MNRAS.428.1116R}
\bibinfo{author}{\bibfnamefont{A.~J.} \bibnamefont{{Ross}}}, \bibinfo{author}{\bibfnamefont{W.~J.} \bibnamefont{{Percival}}}, \bibinfo{author}{\bibfnamefont{A.}~\bibnamefont{{Carnero}}}, \bibinfo{author}{\bibfnamefont{G.-b.} \bibnamefont{{Zhao}}}, \bibinfo{author}{\bibfnamefont{M.}~\bibnamefont{{Manera}}}, \bibinfo{author}{\bibfnamefont{A.}~\bibnamefont{{Raccanelli}}}, \bibinfo{author}{\bibfnamefont{E.}~\bibnamefont{{Aubourg}}}, \bibinfo{author}{\bibfnamefont{D.}~\bibnamefont{{Bizyaev}}}, \bibinfo{author}{\bibfnamefont{H.}~\bibnamefont{{Brewington}}}, \bibinfo{author}{\bibfnamefont{J.}~\bibnamefont{{Brinkmann}}}, \bibnamefont{et~al.}, \bibinfo{journal}{\mnras} \textbf{\bibinfo{volume}{428}}, \bibinfo{pages}{1116} (\bibinfo{year}{2013}), \eprint{1208.1491}.

\bibitem[{\citenamefont{{Rezaie} et~al.}(2021)\citenamefont{{Rezaie}, {Ross}, {Seo}, {Mueller}, {Percival}, {Merz}, {Katebi}, {Bunescu}, {Bautista}, {Brownstein} et~al.}}]{2021MNRAS.506.3439R}
\bibinfo{author}{\bibfnamefont{M.}~\bibnamefont{{Rezaie}}}, \bibinfo{author}{\bibfnamefont{A.~J.} \bibnamefont{{Ross}}}, \bibinfo{author}{\bibfnamefont{H.-J.} \bibnamefont{{Seo}}}, \bibinfo{author}{\bibfnamefont{E.-M.} \bibnamefont{{Mueller}}}, \bibinfo{author}{\bibfnamefont{W.~J.} \bibnamefont{{Percival}}}, \bibinfo{author}{\bibfnamefont{G.}~\bibnamefont{{Merz}}}, \bibinfo{author}{\bibfnamefont{R.}~\bibnamefont{{Katebi}}}, \bibinfo{author}{\bibfnamefont{R.~C.} \bibnamefont{{Bunescu}}}, \bibinfo{author}{\bibfnamefont{J.}~\bibnamefont{{Bautista}}}, \bibinfo{author}{\bibfnamefont{J.~R.} \bibnamefont{{Brownstein}}}, \bibnamefont{et~al.}, \bibinfo{journal}{\mnras} \textbf{\bibinfo{volume}{506}}, \bibinfo{pages}{3439} (\bibinfo{year}{2021}), \eprint{2106.13724}.

\bibitem[{\citenamefont{{Kurita} and {Takada}}(2023)}]{2023arXiv230202925K}
\bibinfo{author}{\bibfnamefont{T.}~\bibnamefont{{Kurita}}} \bibnamefont{and} \bibinfo{author}{\bibfnamefont{M.}~\bibnamefont{{Takada}}}, \bibinfo{journal}{\prd} \textbf{\bibinfo{volume}{108}}, \bibinfo{eid}{083533} (\bibinfo{year}{2023}), \eprint{2302.02925}.

\bibitem[{\citenamefont{{Cagliari} et~al.}(2023)\citenamefont{{Cagliari}, {Castorina}, {Bonici}, and {Bianchi}}}]{2023arXiv230915814C}
\bibinfo{author}{\bibfnamefont{M.~S.} \bibnamefont{{Cagliari}}}, \bibinfo{author}{\bibfnamefont{E.}~\bibnamefont{{Castorina}}}, \bibinfo{author}{\bibfnamefont{M.}~\bibnamefont{{Bonici}}}, \bibnamefont{and} \bibinfo{author}{\bibfnamefont{D.}~\bibnamefont{{Bianchi}}}, \bibinfo{journal}{arXiv e-prints} \bibinfo{eid}{arXiv:2309.15814} (\bibinfo{year}{2023}), \eprint{2309.15814}.

\bibitem[{\citenamefont{{Riquelme} et~al.}(2023)\citenamefont{{Riquelme}, {Avila}, {Garc{\'\i}a-Bellido}, {Porredon}, {Ferrero}, {Chan}, {Rosenfeld}, {Camacho}, {Adame}, {Carnero Rosell} et~al.}}]{2023MNRAS.523..603R}
\bibinfo{author}{\bibfnamefont{W.}~\bibnamefont{{Riquelme}}}, \bibinfo{author}{\bibfnamefont{S.}~\bibnamefont{{Avila}}}, \bibinfo{author}{\bibfnamefont{J.}~\bibnamefont{{Garc{\'\i}a-Bellido}}}, \bibinfo{author}{\bibfnamefont{A.}~\bibnamefont{{Porredon}}}, \bibinfo{author}{\bibfnamefont{I.}~\bibnamefont{{Ferrero}}}, \bibinfo{author}{\bibfnamefont{K.~C.} \bibnamefont{{Chan}}}, \bibinfo{author}{\bibfnamefont{R.}~\bibnamefont{{Rosenfeld}}}, \bibinfo{author}{\bibfnamefont{H.}~\bibnamefont{{Camacho}}}, \bibinfo{author}{\bibfnamefont{A.~G.} \bibnamefont{{Adame}}}, \bibinfo{author}{\bibfnamefont{A.}~\bibnamefont{{Carnero Rosell}}}, \bibnamefont{et~al.}, \bibinfo{journal}{\mnras} \textbf{\bibinfo{volume}{523}}, \bibinfo{pages}{603} (\bibinfo{year}{2023}), \eprint{2209.07187}.

\bibitem[{\citenamefont{{Rezaie} et~al.}(2023)\citenamefont{{Rezaie}, {Ross}, {Seo}, {Kong}, {Porredon}, {Samushia}, {Chaussidon}, {Krolewski}, {de Mattia}, {Beutler} et~al.}}]{2023arXiv230701753R}
\bibinfo{author}{\bibfnamefont{M.}~\bibnamefont{{Rezaie}}}, \bibinfo{author}{\bibfnamefont{A.~J.} \bibnamefont{{Ross}}}, \bibinfo{author}{\bibfnamefont{H.-J.} \bibnamefont{{Seo}}}, \bibinfo{author}{\bibfnamefont{H.}~\bibnamefont{{Kong}}}, \bibinfo{author}{\bibfnamefont{A.}~\bibnamefont{{Porredon}}}, \bibinfo{author}{\bibfnamefont{L.}~\bibnamefont{{Samushia}}}, \bibinfo{author}{\bibfnamefont{E.}~\bibnamefont{{Chaussidon}}}, \bibinfo{author}{\bibfnamefont{A.}~\bibnamefont{{Krolewski}}}, \bibinfo{author}{\bibfnamefont{A.}~\bibnamefont{{de Mattia}}}, \bibinfo{author}{\bibfnamefont{F.}~\bibnamefont{{Beutler}}}, \bibnamefont{et~al.}, \bibinfo{journal}{arXiv e-prints} \bibinfo{eid}{arXiv:2307.01753} (\bibinfo{year}{2023}), \eprint{2307.01753}.

\bibitem[{\citenamefont{{Chaussidon} et~al.}(2024)\citenamefont{{Chaussidon}, {Y{\`e}che}, {de Mattia}, {Payerne}, {McDonald}, {Ross}, {Ahlen}, {Bianchi}, {Brooks}, {Burtin} et~al.}}]{2024arXiv241117623C}
\bibinfo{author}{\bibfnamefont{E.}~\bibnamefont{{Chaussidon}}}, \bibinfo{author}{\bibfnamefont{C.}~\bibnamefont{{Y{\`e}che}}}, \bibinfo{author}{\bibfnamefont{A.}~\bibnamefont{{de Mattia}}}, \bibinfo{author}{\bibfnamefont{C.}~\bibnamefont{{Payerne}}}, \bibinfo{author}{\bibfnamefont{P.}~\bibnamefont{{McDonald}}}, \bibinfo{author}{\bibfnamefont{A.~J.} \bibnamefont{{Ross}}}, \bibinfo{author}{\bibfnamefont{S.}~\bibnamefont{{Ahlen}}}, \bibinfo{author}{\bibfnamefont{D.}~\bibnamefont{{Bianchi}}}, \bibinfo{author}{\bibfnamefont{D.}~\bibnamefont{{Brooks}}}, \bibinfo{author}{\bibfnamefont{E.}~\bibnamefont{{Burtin}}}, \bibnamefont{et~al.}, \bibinfo{journal}{arXiv e-prints} \bibinfo{eid}{arXiv:2411.17623} (\bibinfo{year}{2024}), \eprint{2411.17623}.

\bibitem[{\citenamefont{{Biagetti}}(2019)}]{2019Galax...7...71B}
\bibinfo{author}{\bibfnamefont{M.}~\bibnamefont{{Biagetti}}}, \bibinfo{journal}{Galaxies} \textbf{\bibinfo{volume}{7}}, \bibinfo{eid}{71} (\bibinfo{year}{2019}), \eprint{1906.12244}.

\bibitem[{\citenamefont{{Fang} et~al.}(2020)\citenamefont{{Fang}, {Krause}, {Eifler}, and {MacCrann}}}]{2020JCAP...05..010F}
\bibinfo{author}{\bibfnamefont{X.}~\bibnamefont{{Fang}}}, \bibinfo{author}{\bibfnamefont{E.}~\bibnamefont{{Krause}}}, \bibinfo{author}{\bibfnamefont{T.}~\bibnamefont{{Eifler}}}, \bibnamefont{and} \bibinfo{author}{\bibfnamefont{N.}~\bibnamefont{{MacCrann}}}, \bibinfo{journal}{\jcap} \textbf{\bibinfo{volume}{2020}}, \bibinfo{eid}{010} (\bibinfo{year}{2020}), \eprint{1911.11947}.

\bibitem[{\citenamefont{{Abbott} et~al.}(2022)\citenamefont{{Abbott}, {Aguena}, {Alarcon}, {Allam}, {Alves}, {Amon}, {Andrade-Oliveira}, {Annis}, {Avila}, {Bacon} et~al.}}]{2022PhRvD.105b3520A}
\bibinfo{author}{\bibfnamefont{T.~M.~C.} \bibnamefont{{Abbott}}}, \bibinfo{author}{\bibfnamefont{M.}~\bibnamefont{{Aguena}}}, \bibinfo{author}{\bibfnamefont{A.}~\bibnamefont{{Alarcon}}}, \bibinfo{author}{\bibfnamefont{S.}~\bibnamefont{{Allam}}}, \bibinfo{author}{\bibfnamefont{O.}~\bibnamefont{{Alves}}}, \bibinfo{author}{\bibfnamefont{A.}~\bibnamefont{{Amon}}}, \bibinfo{author}{\bibfnamefont{F.}~\bibnamefont{{Andrade-Oliveira}}}, \bibinfo{author}{\bibfnamefont{J.}~\bibnamefont{{Annis}}}, \bibinfo{author}{\bibfnamefont{S.}~\bibnamefont{{Avila}}}, \bibinfo{author}{\bibfnamefont{D.}~\bibnamefont{{Bacon}}}, \bibnamefont{et~al.}, \bibinfo{journal}{\prd} \textbf{\bibinfo{volume}{105}}, \bibinfo{eid}{023520} (\bibinfo{year}{2022}), \eprint{2105.13549}.

\bibitem[{\citenamefont{{Miyatake} et~al.}(2023)\citenamefont{{Miyatake}, {Sugiyama}, {Takada}, {Nishimichi}, {Li}, {Shirasaki}, {More}, {Kobayashi}, {Nishizawa}, {Rau} et~al.}}]{2023PhRvD.108l3517M}
\bibinfo{author}{\bibfnamefont{H.}~\bibnamefont{{Miyatake}}}, \bibinfo{author}{\bibfnamefont{S.}~\bibnamefont{{Sugiyama}}}, \bibinfo{author}{\bibfnamefont{M.}~\bibnamefont{{Takada}}}, \bibinfo{author}{\bibfnamefont{T.}~\bibnamefont{{Nishimichi}}}, \bibinfo{author}{\bibfnamefont{X.}~\bibnamefont{{Li}}}, \bibinfo{author}{\bibfnamefont{M.}~\bibnamefont{{Shirasaki}}}, \bibinfo{author}{\bibfnamefont{S.}~\bibnamefont{{More}}}, \bibinfo{author}{\bibfnamefont{Y.}~\bibnamefont{{Kobayashi}}}, \bibinfo{author}{\bibfnamefont{A.~J.} \bibnamefont{{Nishizawa}}}, \bibinfo{author}{\bibfnamefont{M.~M.} \bibnamefont{{Rau}}}, \bibnamefont{et~al.}, \bibinfo{journal}{\prd} \textbf{\bibinfo{volume}{108}}, \bibinfo{eid}{123517} (\bibinfo{year}{2023}), \eprint{2304.00704}.

\bibitem[{\citenamefont{{Terasawa} et~al.}(2024)\citenamefont{{Terasawa}, {Li}, {Takada}, {Nishimichi}, {Tanaka}, {Sugiyama}, {Kurita}, {Zhang}, {Shirasaki}, {Takahashi} et~al.}}]{Terasawa24}
\bibinfo{author}{\bibfnamefont{R.}~\bibnamefont{{Terasawa}}}, \bibinfo{author}{\bibfnamefont{X.}~\bibnamefont{{Li}}}, \bibinfo{author}{\bibfnamefont{M.}~\bibnamefont{{Takada}}}, \bibinfo{author}{\bibfnamefont{T.}~\bibnamefont{{Nishimichi}}}, \bibinfo{author}{\bibfnamefont{S.}~\bibnamefont{{Tanaka}}}, \bibinfo{author}{\bibfnamefont{S.}~\bibnamefont{{Sugiyama}}}, \bibinfo{author}{\bibfnamefont{T.}~\bibnamefont{{Kurita}}}, \bibinfo{author}{\bibfnamefont{T.}~\bibnamefont{{Zhang}}}, \bibinfo{author}{\bibfnamefont{M.}~\bibnamefont{{Shirasaki}}}, \bibinfo{author}{\bibfnamefont{R.}~\bibnamefont{{Takahashi}}}, \bibnamefont{et~al.}, \bibinfo{journal}{arXiv e-prints} \bibinfo{eid}{arXiv:2403.20323} (\bibinfo{year}{2024}), \eprint{2403.20323}.

\bibitem[{\citenamefont{{Jeong} et~al.}(2012)\citenamefont{{Jeong}, {Schmidt}, and {Hirata}}}]{2012PhRvD..85b3504J}
\bibinfo{author}{\bibfnamefont{D.}~\bibnamefont{{Jeong}}}, \bibinfo{author}{\bibfnamefont{F.}~\bibnamefont{{Schmidt}}}, \bibnamefont{and} \bibinfo{author}{\bibfnamefont{C.~M.} \bibnamefont{{Hirata}}}, \bibinfo{journal}{\prd} \textbf{\bibinfo{volume}{85}}, \bibinfo{eid}{023504} (\bibinfo{year}{2012}), \eprint{1107.5427}.

\bibitem[{\citenamefont{{Planck Collaboration} et~al.}(2016)\citenamefont{{Planck Collaboration}, {Ade}, {Aghanim}, {Arnaud}, {Ashdown}, {Aumont}, {Baccigalupi}, {Banday}, {Barreiro}, {Bartlett} et~al.}}]{planck-collaboration:2015fj}
\bibinfo{author}{\bibnamefont{{Planck Collaboration}}}, \bibinfo{author}{\bibfnamefont{P.~A.~R.} \bibnamefont{{Ade}}}, \bibinfo{author}{\bibfnamefont{N.}~\bibnamefont{{Aghanim}}}, \bibinfo{author}{\bibfnamefont{M.}~\bibnamefont{{Arnaud}}}, \bibinfo{author}{\bibfnamefont{M.}~\bibnamefont{{Ashdown}}}, \bibinfo{author}{\bibfnamefont{J.}~\bibnamefont{{Aumont}}}, \bibinfo{author}{\bibfnamefont{C.}~\bibnamefont{{Baccigalupi}}}, \bibinfo{author}{\bibfnamefont{A.~J.} \bibnamefont{{Banday}}}, \bibinfo{author}{\bibfnamefont{R.~B.} \bibnamefont{{Barreiro}}}, \bibinfo{author}{\bibfnamefont{J.~G.} \bibnamefont{{Bartlett}}}, \bibnamefont{et~al.}, \bibinfo{journal}{\aap} \textbf{\bibinfo{volume}{594}}, \bibinfo{eid}{A13} (\bibinfo{year}{2016}), \eprint{1502.01589}.

\bibitem[{\citenamefont{Collaboration et~al.}(2021)\citenamefont{Collaboration, Mandelbaum, Eifler, Hložek, Collett, Gawiser, Scolnic, Alonso, Awan, Biswas et~al.}}]{thelsstdarkenergysciencecollaboration2021lsst}
\bibinfo{author}{\bibfnamefont{T.~L. D. E.~S.} \bibnamefont{Collaboration}}, \bibinfo{author}{\bibfnamefont{R.}~\bibnamefont{Mandelbaum}}, \bibinfo{author}{\bibfnamefont{T.}~\bibnamefont{Eifler}}, \bibinfo{author}{\bibfnamefont{R.}~\bibnamefont{Hložek}}, \bibinfo{author}{\bibfnamefont{T.}~\bibnamefont{Collett}}, \bibinfo{author}{\bibfnamefont{E.}~\bibnamefont{Gawiser}}, \bibinfo{author}{\bibfnamefont{D.}~\bibnamefont{Scolnic}}, \bibinfo{author}{\bibfnamefont{D.}~\bibnamefont{Alonso}}, \bibinfo{author}{\bibfnamefont{H.}~\bibnamefont{Awan}}, \bibinfo{author}{\bibfnamefont{R.}~\bibnamefont{Biswas}}, \bibnamefont{et~al.}, \emph{\bibinfo{title}{The lsst dark energy science collaboration (desc) science requirements document}} (\bibinfo{year}{2021}), \eprint{1809.01669}.

\bibitem[{\citenamefont{{Schmittfull} and {Seljak}}(2018)}]{2018PhRvD..97l3540S}
\bibinfo{author}{\bibfnamefont{M.}~\bibnamefont{{Schmittfull}}} \bibnamefont{and} \bibinfo{author}{\bibfnamefont{U.}~\bibnamefont{{Seljak}}}, \bibinfo{journal}{\prd} \textbf{\bibinfo{volume}{97}}, \bibinfo{eid}{123540} (\bibinfo{year}{2018}), \eprint{1710.09465}.

\bibitem[{\citenamefont{{Harikane} et~al.}(2022)\citenamefont{{Harikane}, {Ono}, {Ouchi}, {Liu}, {Sawicki}, {Shibuya}, {Behroozi}, {He}, {Shimasaku}, {Arnouts} et~al.}}]{GOLDRUSH_IV}
\bibinfo{author}{\bibfnamefont{Y.}~\bibnamefont{{Harikane}}}, \bibinfo{author}{\bibfnamefont{Y.}~\bibnamefont{{Ono}}}, \bibinfo{author}{\bibfnamefont{M.}~\bibnamefont{{Ouchi}}}, \bibinfo{author}{\bibfnamefont{C.}~\bibnamefont{{Liu}}}, \bibinfo{author}{\bibfnamefont{M.}~\bibnamefont{{Sawicki}}}, \bibinfo{author}{\bibfnamefont{T.}~\bibnamefont{{Shibuya}}}, \bibinfo{author}{\bibfnamefont{P.~S.} \bibnamefont{{Behroozi}}}, \bibinfo{author}{\bibfnamefont{W.}~\bibnamefont{{He}}}, \bibinfo{author}{\bibfnamefont{K.}~\bibnamefont{{Shimasaku}}}, \bibinfo{author}{\bibfnamefont{S.}~\bibnamefont{{Arnouts}}}, \bibnamefont{et~al.}, \bibinfo{journal}{\apjs} \textbf{\bibinfo{volume}{259}}, \bibinfo{eid}{20} (\bibinfo{year}{2022}), \eprint{2108.01090}.

\bibitem[{\citenamefont{{Crocce} et~al.}(2011)\citenamefont{{Crocce}, {Cabr{\'e}}, and {Gazta{\~n}aga}}}]{2011MNRAS.414..329C}
\bibinfo{author}{\bibfnamefont{M.}~\bibnamefont{{Crocce}}}, \bibinfo{author}{\bibfnamefont{A.}~\bibnamefont{{Cabr{\'e}}}}, \bibnamefont{and} \bibinfo{author}{\bibfnamefont{E.}~\bibnamefont{{Gazta{\~n}aga}}}, \bibinfo{journal}{\mnras} \textbf{\bibinfo{volume}{414}}, \bibinfo{pages}{329} (\bibinfo{year}{2011}), \eprint{1004.4640}.

\bibitem[{\citenamefont{{Takada} and {Jain}}(2003)}]{2003MNRAS.344..857T}
\bibinfo{author}{\bibfnamefont{M.}~\bibnamefont{{Takada}}} \bibnamefont{and} \bibinfo{author}{\bibfnamefont{B.}~\bibnamefont{{Jain}}}, \bibinfo{journal}{\mnras} \textbf{\bibinfo{volume}{344}}, \bibinfo{pages}{857} (\bibinfo{year}{2003}), \eprint{astro-ph/0304034}.

\bibitem[{\citenamefont{{Joachimi} et~al.}(2008)\citenamefont{{Joachimi}, {Schneider}, and {Eifler}}}]{2008A&A...477...43J}
\bibinfo{author}{\bibfnamefont{B.}~\bibnamefont{{Joachimi}}}, \bibinfo{author}{\bibfnamefont{P.}~\bibnamefont{{Schneider}}}, \bibnamefont{and} \bibinfo{author}{\bibfnamefont{T.}~\bibnamefont{{Eifler}}}, \bibinfo{journal}{\aap} \textbf{\bibinfo{volume}{477}}, \bibinfo{pages}{43} (\bibinfo{year}{2008}), \eprint{0708.0387}.

\bibitem[{\citenamefont{Senatore and Zaldarriaga}(2015)}]{Senatore_2015}
\bibinfo{author}{\bibfnamefont{L.}~\bibnamefont{Senatore}} \bibnamefont{and} \bibinfo{author}{\bibfnamefont{M.}~\bibnamefont{Zaldarriaga}}, \bibinfo{journal}{Journal of Cosmology and Astroparticle Physics} \textbf{\bibinfo{volume}{2015}}, \bibinfo{pages}{013–013} (\bibinfo{year}{2015}), ISSN \bibinfo{issn}{1475-7516}, \urlprefix\url{http://dx.doi.org/10.1088/1475-7516/2015/02/013}.

\bibitem[{\citenamefont{{Ellis} and {Baldwin}}(1984)}]{1984MNRAS.206..377E}
\bibinfo{author}{\bibfnamefont{G.~F.~R.} \bibnamefont{{Ellis}}} \bibnamefont{and} \bibinfo{author}{\bibfnamefont{J.~E.} \bibnamefont{{Baldwin}}}, \bibinfo{journal}{\mnras} \textbf{\bibinfo{volume}{206}}, \bibinfo{pages}{377} (\bibinfo{year}{1984}).

\bibitem[{\citenamefont{{Blake} and {Wall}}(2002)}]{2002Natur.416..150B}
\bibinfo{author}{\bibfnamefont{C.}~\bibnamefont{{Blake}}} \bibnamefont{and} \bibinfo{author}{\bibfnamefont{J.}~\bibnamefont{{Wall}}}, \bibinfo{journal}{\nat} \textbf{\bibinfo{volume}{416}}, \bibinfo{pages}{150} (\bibinfo{year}{2002}), \eprint{astro-ph/0203385}.

\bibitem[{\citenamefont{{Itoh} et~al.}(2010)\citenamefont{{Itoh}, {Yahata}, and {Takada}}}]{2010PhRvD..82d3530I}
\bibinfo{author}{\bibfnamefont{Y.}~\bibnamefont{{Itoh}}}, \bibinfo{author}{\bibfnamefont{K.}~\bibnamefont{{Yahata}}}, \bibnamefont{and} \bibinfo{author}{\bibfnamefont{M.}~\bibnamefont{{Takada}}}, \bibinfo{journal}{\prd} \textbf{\bibinfo{volume}{82}}, \bibinfo{eid}{043530} (\bibinfo{year}{2010}), \eprint{0912.1460}.

\bibitem[{\citenamefont{{Takada} et~al.}(2014)\citenamefont{{Takada}, {Ellis}, {Chiba}, {Greene}, {Aihara}, {Arimoto}, {Bundy}, {Cohen}, {Dor{\'e}}, {Graves} et~al.}}]{2014PASJ...66R...1T}
\bibinfo{author}{\bibfnamefont{M.}~\bibnamefont{{Takada}}}, \bibinfo{author}{\bibfnamefont{R.~S.} \bibnamefont{{Ellis}}}, \bibinfo{author}{\bibfnamefont{M.}~\bibnamefont{{Chiba}}}, \bibinfo{author}{\bibfnamefont{J.~E.} \bibnamefont{{Greene}}}, \bibinfo{author}{\bibfnamefont{H.}~\bibnamefont{{Aihara}}}, \bibinfo{author}{\bibfnamefont{N.}~\bibnamefont{{Arimoto}}}, \bibinfo{author}{\bibfnamefont{K.}~\bibnamefont{{Bundy}}}, \bibinfo{author}{\bibfnamefont{J.}~\bibnamefont{{Cohen}}}, \bibinfo{author}{\bibfnamefont{O.}~\bibnamefont{{Dor{\'e}}}}, \bibinfo{author}{\bibfnamefont{G.}~\bibnamefont{{Graves}}}, \bibnamefont{et~al.}, \bibinfo{journal}{\pasj} \textbf{\bibinfo{volume}{66}}, \bibinfo{eid}{R1} (\bibinfo{year}{2014}), \eprint{1206.0737}.

\bibitem[{\citenamefont{Hand et~al.}(2018)\citenamefont{Hand, Feng, Beutler, Li, Modi, Seljak, and Slepian}}]{Hand_2018}
\bibinfo{author}{\bibfnamefont{N.}~\bibnamefont{Hand}}, \bibinfo{author}{\bibfnamefont{Y.}~\bibnamefont{Feng}}, \bibinfo{author}{\bibfnamefont{F.}~\bibnamefont{Beutler}}, \bibinfo{author}{\bibfnamefont{Y.}~\bibnamefont{Li}}, \bibinfo{author}{\bibfnamefont{C.}~\bibnamefont{Modi}}, \bibinfo{author}{\bibfnamefont{U.}~\bibnamefont{Seljak}}, \bibnamefont{and} \bibinfo{author}{\bibfnamefont{Z.}~\bibnamefont{Slepian}}, \bibinfo{journal}{The Astronomical Journal} \textbf{\bibinfo{volume}{156}}, \bibinfo{pages}{160} (\bibinfo{year}{2018}), ISSN \bibinfo{issn}{1538-3881}, \urlprefix\url{http://dx.doi.org/10.3847/1538-3881/aadae0}.

\end{thebibliography}

\end{document}